%% file: mmw03_bulge.tex
\documentclass{aa}  

\input{input}

\begin{document} 

\title{\papername\ III. Panoramic view of the bulge}

\titlerunning{Rediscovering the Milky Way bulge}

\authorrunning{S. Khoperskov et al.}

\author{Sergey Khoperskov$^1$\thanks{sergey.khoperskov@gmail.com}\orcidlink{0000-0003-2105-0763}, 
Paola Di Matteo$^2$, 
Matthias Steinmetz$^1$\orcidlink{0000-0001-6516-7459}, 
Bridget Ratcliffe$^1$\orcidlink{0000-0003-1124-7378},
Glenn van de Ven$^3$\orcidlink{0000-0003-4546-7731},  \\
Tristan Boin$^2$\orcidlink{0009-0001-6758-9855}, 
Misha Haywood$^2$\orcidlink{0000-0003-0434-0400}, 
Nikolay Kacharov$^1$\orcidlink{0000-0002-6072-6669}, 
Ivan Minchev$^1$\orcidlink{0000-0002-5627-0355}, \\
Davor Krajnovic$^1$,
Marica Valentini$^1$\orcidlink{0000-0003-0974-4148}, 
Roelof S. de Jong$^1$\orcidlink{0000-0001-6982-4081} }

\institute{$^1$ Leibniz-Institut für Astrophysik Potsdam (AIP),
              An der Sternwarte 16, 14482 Potsdam, Germany \\
              $^2$ GEPI, Observatoire de Paris, PSL Research University, CNRS, Place Jules Janssen, 92195 Meudon, France \\
              $^3$ Department of Astrophysics, University of Vienna, Türkenschanzstraße 17, A-1180 Vienna, Austria 
              }

\abstract{The innermost parts of the Milky Way~(MW) are very difficult to observe due to the high extinction along the line of sight, especially close to the disc mid-plane. However, this region contains the most massive complex stellar component of the MW, the bulge, primarily composed of disc stars whose structure is (re-)shaped by the evolution of the bar. In this work, we extend the application of the orbit superposition method to explore the present-day 3D structure, orbital composition, chemical abundance trends and kinematics of the MW bulge. 

Thanks to our approach, we are able to transfer astrometry from Gaia and stellar parameters from APOGEE DR 17 to map the inner MW without obscuration by the survey footprint and selection function. We demonstrate that the MW bulge is made of two main populations originating from a metal-poor, high-$\alpha$ thick disc and a metal-rich, low-$\alpha$ thin disc, with a mass ratio of 4:3, seen as two major components in the metallicity distribution function~(MDF). Finer MDF structures hint at multiple sub-populations associated with different orbital families of the bulge, which, however, have broad MDFs themselves. Decomposition using 2D Gaussian Mixture Models in the \FeH-\MgFe plane identifies five components, including a population with ex-situ origin. Two dominant ones correspond to the thin and thick discs and two in between trace the transition between them. We show that no universal metallicity gradient value can characterise the MW bulge. The radial gradients closely trace the X-shaped bulge density structure, while the vertical gradient variations follow the boxy component. While having, on average, subsolar metallicity, the MW bulge populations are more metal-rich compared to the surrounding disc, in agreement with extragalactic observations and state-of-the-art simulations reinforcing its secular origin.
}
   

\maketitle


\section{Introduction}

\begin{figure*}
    \centering
    \includegraphics[width=1\hsize]{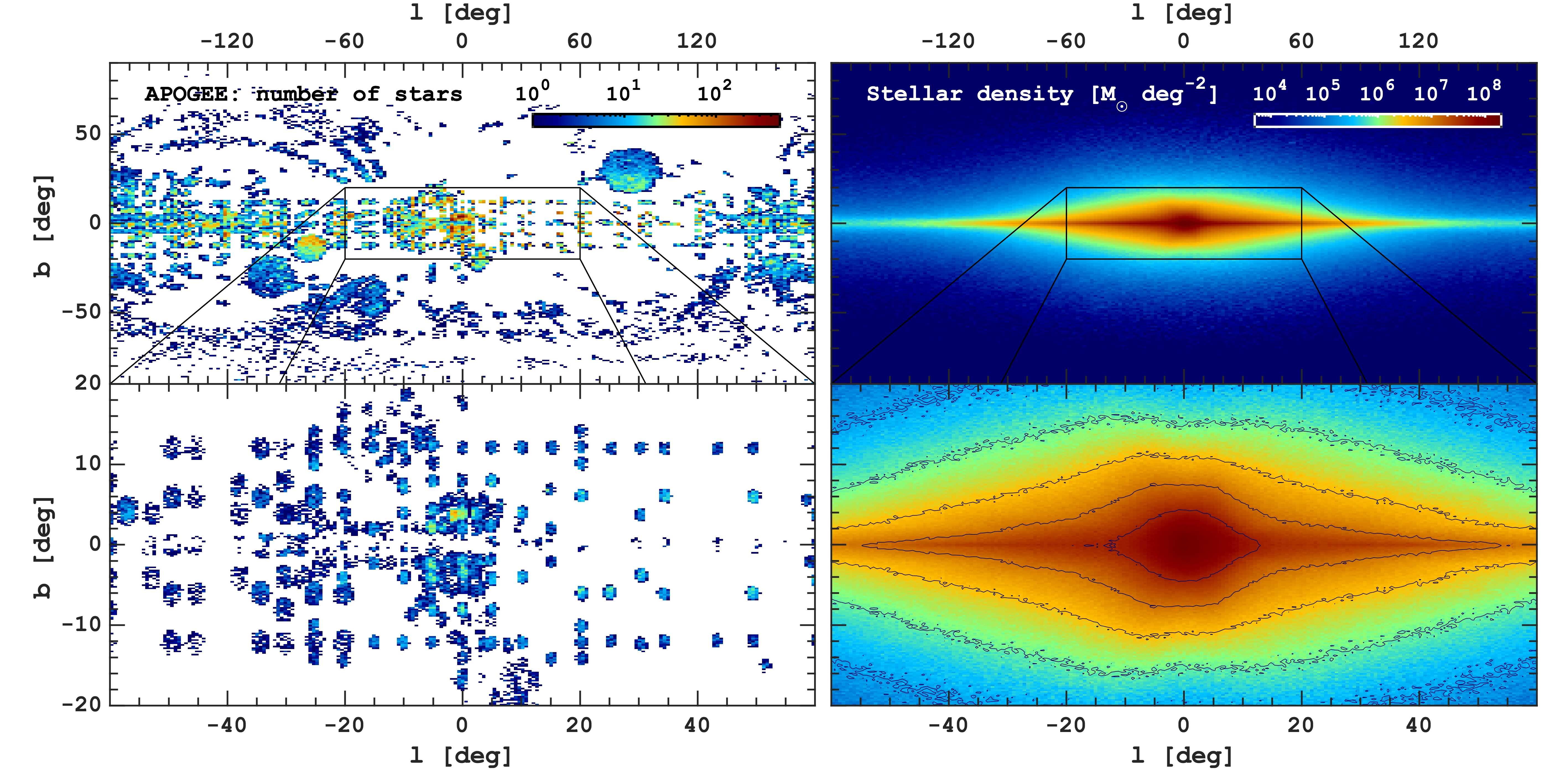}
    \caption{On sky distribution of the APOGEE stars in our initial sample~(left) and stellar mass-weighted density projection obtained by superposition of orbits of these stars~(right). The top panels cover the entire disc of the MW, while the bottom ones zoom in into the MW bulge region. The orbit superposition technique not only reconstructs the 3D density distribution of the MW disc and bulge but also maps the abundance patterns of APOGEE stars across these components, effectively eliminating the limitations caused by the survey’s footprint and correcting for selection function biases. The reconstructed density shows a prominent boxy/X-shaped bulge structure, which is asymmetric along the longitude due to the bar orientation.}
    \label{fig03::bulge_den_vs_APOGEE}
\end{figure*}

\begin{figure}
    \centering
    \includegraphics[width=1\hsize]{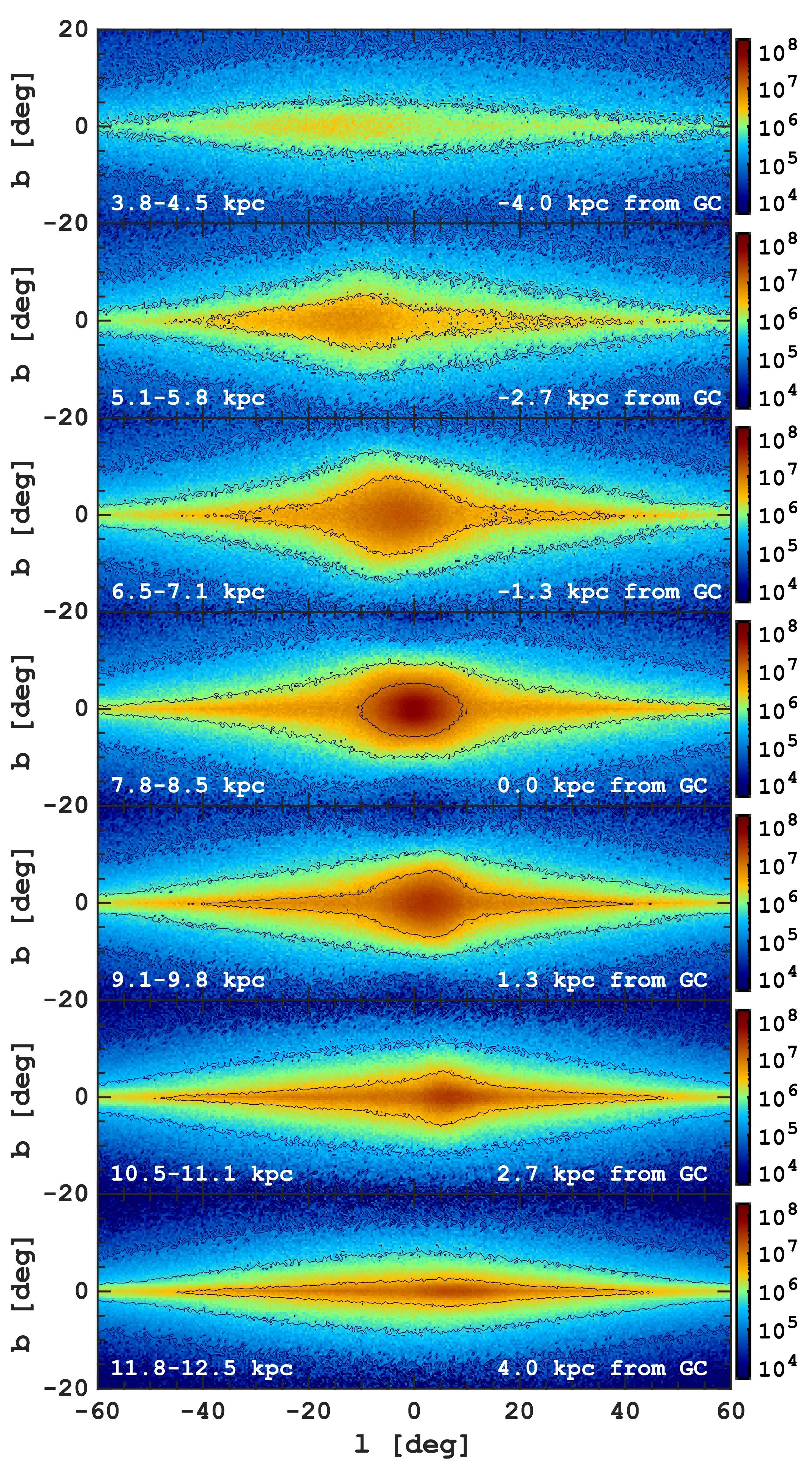}
    \caption{Variation of the inner MW density structure along the line-of-sight reconstructed using orbit superposition and APOGEE data. From top to bottom, the panels show the stellar density in (l,b) coordinates in $0.6$~kpc-width slices with increasing distances from the Sun, as marked in the bottom left of each panel. The middle panel shows the boxy bulge structure at the Galactic centre, while the upper and lower panels show the near and far sides of the bar/bulge, respectively. As we slice the MW bar inclined relative to the Sun - Galactic centre line, the off-centred X-shaped structure transitions from negative to positive longitudes with increasing distance.}
    \label{fig03::bulge_den_2D_structure}
\end{figure}

\begin{figure*}
    \centering
    \includegraphics[width=1\hsize]{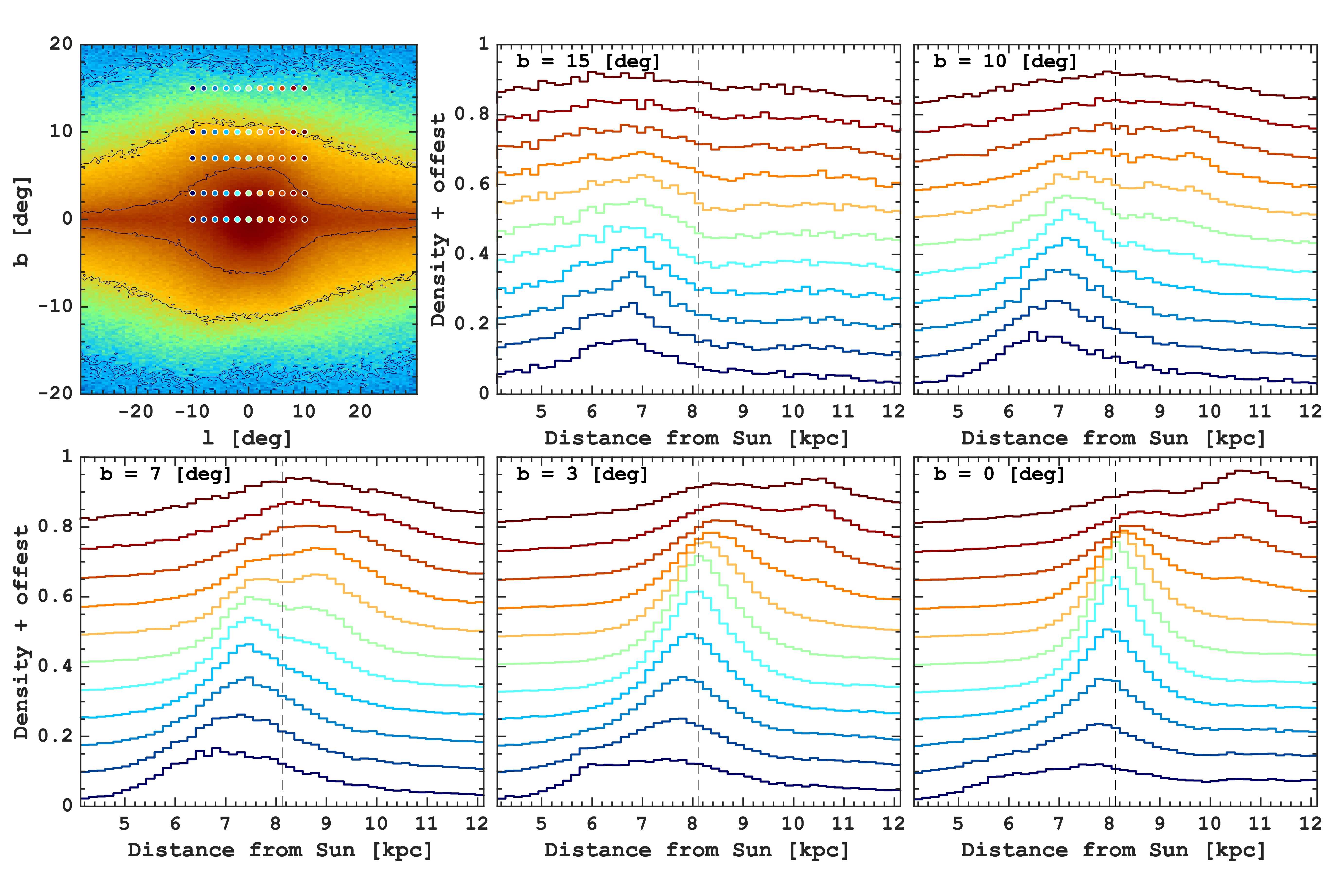}
    \caption{Line-of-sight density structure of the MW bulge reconstructed using orbit superposition and APOGEE data. The top left panel shows the selection of the bulge fields marked by circles of different colour with the total column stellar density on the background. Other panels show the stellar density distribution along the line-of-sight at different latitudes, as marked in the top left panel. The colour of the lines matches the colour of the circles in the top left panel. Only fields with $b\geq0$ are shown as the data are symmetric relative to the midplane.  \href{https://www.dropbox.com/scl/fi/b70pu935bufjfd6rg5vg5/evolution_LOS_bulge_structure.mp4?rlkey=buj6dlzj1vfyvsuxr0r731rab&dl=0}{Animation}}
    \label{fig03::bulge_den_1D_structure}
\end{figure*}

\begin{figure}
    \centering
    \includegraphics[width=1\hsize]{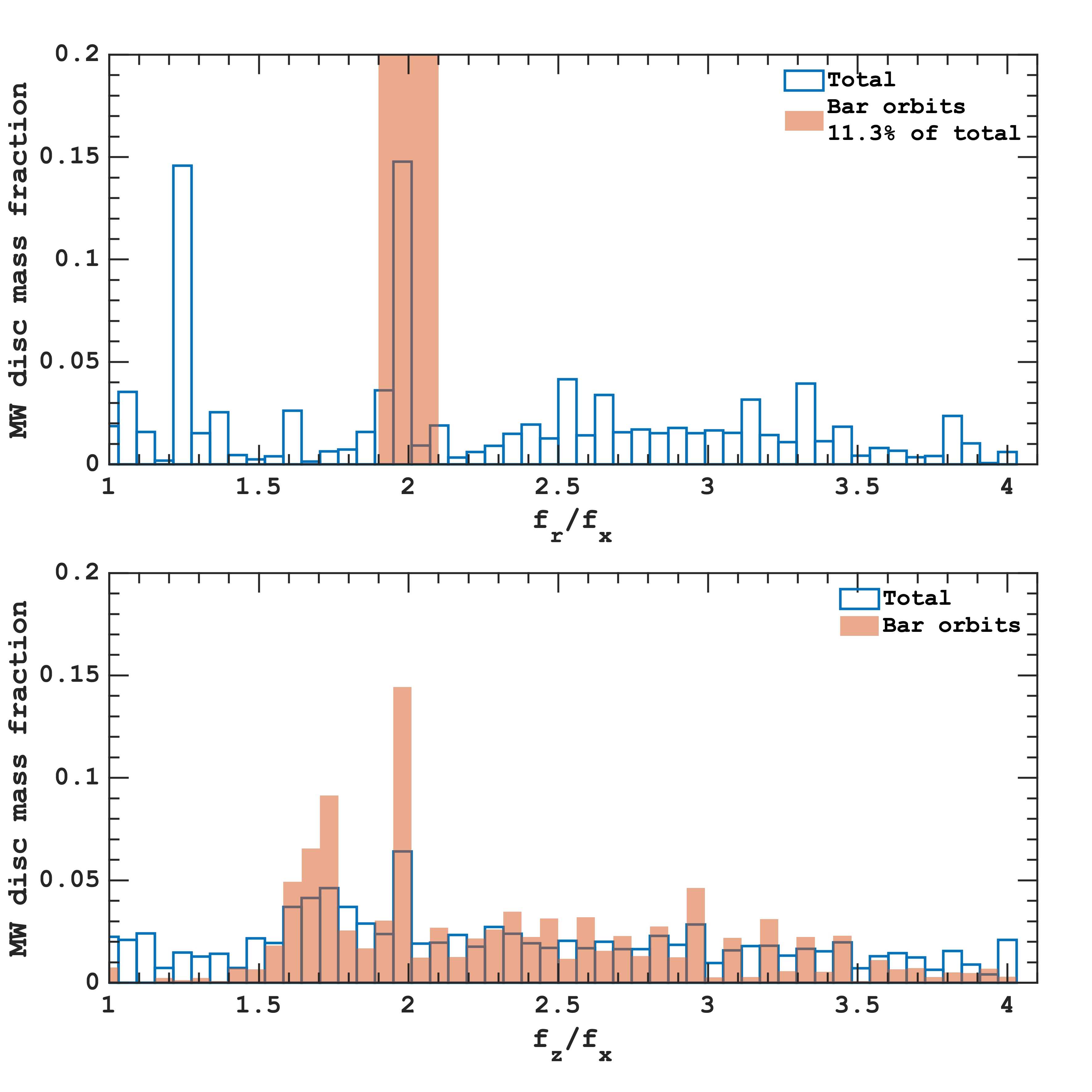}
    \caption{Orbital frequency distribution of the MW bulge populations. The top panel shows the stellar mass-weighted distribution of the in-plane orbital frequency ratio, $f_r/f_x$. The elongated orbits, $f_r/f_x \approx 2$, correspond to the bar, while the rest do not support the bar and represent the disc. The bottom panel shows the distribution of the vertical-to-radial orbital frequencies, $f_z/f_x$. Frequency ratio $f_z/f_x\approx 2$ corresponds to the banana-like orbits. The lower values~($f_z/f_x\in1.4-2$) are pretzel-like populations, dominating in the X-shaped bulge structure, whose extent along the bar major axis decreases with decreasing frequency ratio. Orbits with $f_z/f_x > 2$ correspond to the longer bar component~($>2.5-3$kpc) outside the bulge.}
    \label{fig03::bulge_freq_distr}
\end{figure}

\begin{figure*}
    \centering
    \includegraphics[width=1\hsize]{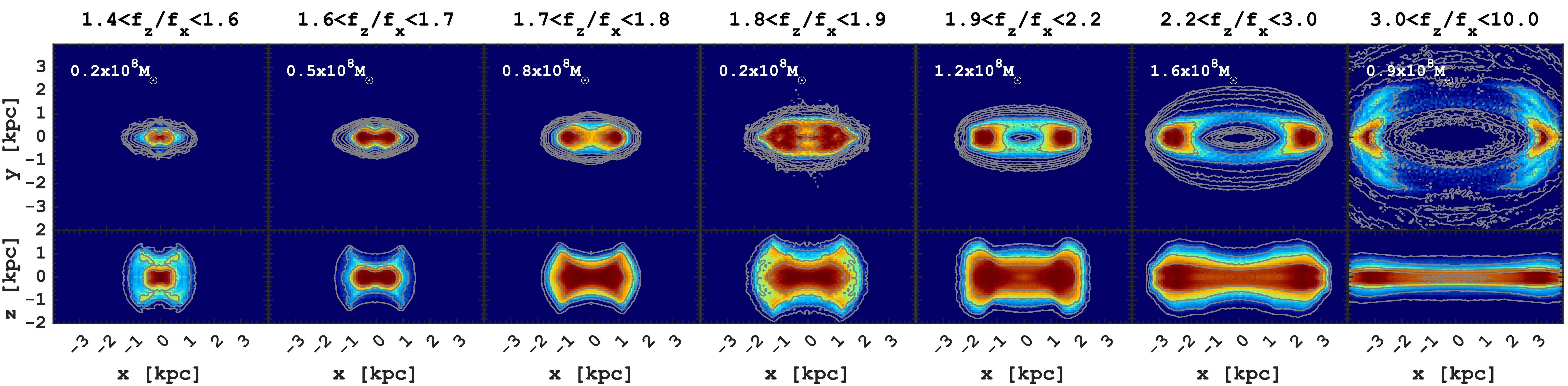}
    \caption{Orbital decomposition of the MW bulge in face-on (top) and side-on (bottom) projections. The panels show the projected stellar density of orbits classified by their vertical-to-in-plane frequency ratio, $f_z/f_x$ (see Fig.~\ref{fig03::bulge_freq_distr}). The bar’s major axis is aligned along the X-axis in both projections. Each orbital class represents a distinct dynamical population, and the stellar mass associated with each class is indicated in the top left corner of the corresponding panels. }
    \label{fig03::bulge_orbits}
\end{figure*}

Although the exact mechanism is still debated~\citep{2020MNRAS.495.3175S}, nowadays, there is a general consensus that the bulge of the MW is the product of the thickening of the bar~\citep[see, e.g.][]{1981A&A....96..164C, 1991Natur.352..411R, 2008IAUS..245...93A, 2005ApJ...628..678D, 2004ApJ...604L..93D,2006ApJ...645..209D, 2010ApJ...720L..72S, 2011ApJ...734L..20M}. Since galactic bars are a natural outcome of the disc instability~\citep{1973ApJ...186..467O, 1981A&A....96..164C,1981A&A....99..362S, 1982MNRAS.199.1069E, 1993RPPh...56..173S}, the MW bulge is predominantly made of the disc stars and, thus, inherits their age and abundance composition ``scrambled'' during formation and secular evolution of the bar~\citep{2010A&A...519A..77B, 2007ApJ...665L..31S,1994AJ....108.2154Z,2006A&A...457L...1Z, 2010A&A...512A..41B, 2012ApJ...749..175J, 2013MNRAS.432.2092N, 2014ApJ...787L..19N, 2015MNRAS.450.4050W, 2016PASA...33...40M, 2016A&A...593A..82H, 2016PASA...33...27D, 2017PASA...34...41N, 2017A&A...607L...4F, 2020MNRAS.494.5936F, 2012ApJ...749..175J, 2022MNRAS.515.1469J}. There is no reason to believe that the boxy/peanut structures observed in external galaxies have experienced different evolutionary histories~\citep{1999AJ....118..126B, 2017MNRAS.468.2058E}. Hence, the MW bulge serves as a probe for understanding general concepts of the secular evolution of barred galaxies~\citep{2004ARA&A..42..603K, 2011MNRAS.415.3308G,2014RvMP...86....1S}.

Large-scale near-infrared observations first unveiled the peculiar morphology of the MW bulge~\citep{1977Natur.265..515O, 1995ApJ...445..716D}, a feature that was further refined with mid-infrared WISE data, revealing its distinct X-shaped residual structure~\citep{2016AJ....152...14N}. The extensive analysis of the bulge using VVV red clump stars~\citep{2010NewA...15..433M, 2018MNRAS.474.1826S} has made possible the reconstruction of a complete 3D mass~(baryons and dark) distribution~\citep{2013MNRAS.435.1874W, 2015MNRAS.448..713P, 2017MNRAS.465.1621P} and kinematics~\citep{2019MNRAS.489.3519C,2019MNRAS.487.5188S} in great detail. In particular, the density distribution of the Galactic bulge is characteristic of a strongly boxy/peanut-shaped bulge within a barred galaxy qualitatively very similar to that predicted by $N$-body simulations of the buckling instability of the disc~\citep{1991Natur.352..411R, 2002MNRAS.330...35A,  2005ApJ...628..678D, 2006ApJ...637..214M, 2014MNRAS.437.1284Q}. It turned out that the chemical abundance patterns of the bulge stars can be found in populations of stars located at the solar radius~\citep{2010A&A...513A..35A, 2010A&A...512A..41B,2017A&A...605A..89B, 2010A&A...509A..20R,2011A&A...530A..54G,2015A&A...584A..46G, 2016PASA...33...22N,2016A&A...593A..82H, 2017A&A...605A..89B, 2020A&A...634A.130B, 2021A&A...655A.117B, 2024ApJ...964...96N}, suggesting that the bulge is composed of disc populations that extend to the Sun. Although the disc of the MW near the solar radius is a superposition of geometric thick and thin components or $\alpha$-rich and $\alpha$-poor populations, respectively~\citep{2003A&A...410..527B, 2003MNRAS.340..304R, 2004A&A...418..989N, 2009A&A...501..941H, 2012ApJ...753..148B, 2013A&A...560A.109H, 2013A&A...558A...9M}, the distinction between geometrically and chemically defined disc components has no trivial solution within the inner MW region and the individual contribution of different~(chemically and geometrically-defined) disc components, can not be easily constrained, partially because of the lack of panoramic coverage of the bulge by spectroscopic data. 

Observational data and models tailored to match the MW bar and bulge show that stellar populations with different kinematics are mapped differently into these components~\citep{1990ApJ...362..604R,1995MNRAS.277.1293M, 2007ApJ...665L..31S,2010A&A...519A..77B, 2011A&A...534A..80H, 2011MNRAS.416L..60B, 2012AJ....143...57K, 2013MNRAS.432.2092N, 2013MNRAS.430..836N, 2016PASA...33...27D, 2017A&A...599A..12Z, 2021A&A...653A.143W}. \cite{2017MNRAS.469.1587D} introduced this effect as kinematic fractionation, which was also observed in isolated~\citep{2017A&A...607L...4F,2019A&A...628A..11D} and self-consistent galaxy-formation simulations~\citep{2018ApJ...861...88B, 2020MNRAS.494.5936F}. According to this mechanism, stars with hot kinematics, before the emergence of the bulge, show a less prominent boxy-peanut structure, while the coldest populations reveal the sharpest boxy-peanut morphology. This phenomenon is seen in the MW, for example, in the magnitude distribution of red giant clump stars in the MW~\citep{2010ApJ...721L..28N, 2010ApJ...724.1491M, 2012ApJ...756...22N, 2014A&A...569A.103R}, where, in particular, for kinematically hot populations, the two lobes of the peanut appear at higher latitudes compared to the colder populations. This, however, requires association of kinematics with chemistry~(or geometry) of stellar populations, which is not so straightforward because some simulations~\citep[see, e.g.][]{2017A&A...606A..47F} suggest that in the outer parts of the boxy/peanut, the hot population tends to have a higher line-of-sight velocity, while in the inner regions, this trend can be reversed, depending also on the orientation of the bar with respect to the line of sight. Such a picture seems to be confirmed observationally by \cite{2012ApJ...756...22N} and \cite{2014A&A...569A.103R} using ARGOS and GES data, respectively, but not so certain in \cite{2016ApJ...819....2N} with the APOGEE data.

Since early studies, there was no consensus on whether the bulge is dominated by the metal-rich~\citep{1988AJ.....95..828R, 1983ApJ...274..723W, 2007ApJ...665L.119R,2015A&A...584A..45S} or old metal-poor stellar populations~\citep{1995Natur.377..701O, 2000A&A...355..949F, 2003A&A...399..931Z}. With the arrival of data from massive spectroscopic surveys, the detailed metallicity distribution function~(MDF) exploration of the bulge becomes possible~\citep{2003A&A...399..931Z, 2008A&A...486..177Z,2013A&A...549A.147B, 2013MNRAS.430..836N, 2017A&A...605A..89B,2017A&A...600A..14S}. Traditionally, the bulge MDF is described as a mixture of several Gaussian components with varying contributions depending on the field location~\citep{2013MNRAS.430..836N,2014A&A...569A.103R, 2017A&A...601A.140R}. Using data from the Gaia-ESO~\cite{2017A&A...599A..12Z} proposed a bi-modal bulge MDF~\citep[see, also][]{2010A&A...519A..77B, 2011A&A...534A..80H, 2013A&A...552A.110G, 2014A&A...569A.103R, 2017A&A...601A.140R, 2017A&A...599A..12Z, 2017A&A...600A..14S, 2018A&A...618A.147Z, 2020AJ....159..270K}. \cite{2020MNRAS.499.1037R} identified three MDF components in the bulge using the APOGEE DR16 data, while \cite{2021A&A...653A.143W} find that the bulge is best fit by four Gaussians where the most metal-poor one is not very significant, thus agreeing with the three dominant components. Using data sets that differ completely in size and spatial coverage, \cite{2013MNRAS.430..836N}~(ARGOS) and \cite{2013A&A...549A.147B, 2017A&A...605A..89B}~(microlensed dwarf and subgiant stars) identified five bulge MDF components whose locations match each other reasonably well. These components were tentatively associated with different stellar populations spanning the metallicity range from $\approx-2$ to $\approx0.5$: a thin boxy/peanut-shaped bulge, a thicker boxy peanut-shaped bulge, the pre-instability thick disc, the metal-weak thick disc and the stellar halo. Nevertheless, all the above-mentioned studies agree that the MDF varies significantly depending on the position in the bulge region. Also, the resulting MDF may be affected by the intrinsic properties of a given survey, e.g. metallicity scale, selection function and the survey footprint~\citep{2021A&A...653A.143W}. The latter is extremely patchy for spectroscopic surveys as the bulge stars are very hard to observe due to high extinction towards the Galactic centre, especially close to the mid-plane. Therefore, obtaining unbiased spectroscopic data for the bulge region is challenging and remains the focus of several future high-resolution MW disc surveys like 4MOST~\citep{2019Msngr.175...35B}, SDSSV~\citep{2023ApJS..267...44A} and especially MOONS~\citep{2020Msngr.180...18G}, which will deliver high-signal-to-noise spectra in the near-infrared~\citep{2020Msngr.180...10C}. Therefore, so far, the question of whether the bulge is a metal-rich or metal-poor component of the MW galaxy remains disputable~\citep{2020MNRAS.499.1037R, 2020MNRAS.494.5936F}.

It is believed that the bulge contains a (very) metal-poor component associated with a spheroidal or classical bulge, characterized by its lack of rotation~\citep{2008A&A...486..177Z, 2013ApJ...776L..19D, 2016ApJ...821L..25K, 2020A&A...641A..96S, 2020MNRAS.491L..11A, 2023MNRAS.526.2365B}. However, the mass contribution of this component appears to be quite small, accounting for only a few percent of the total bulge mass~\citep{2010ApJ...720L..72S, 2012AJ....143...57K, 2013MNRAS.430..836N, 2016ApJ...821L..25K, 2014A&A...567A.122D, 2017MNRAS.469.1587D}. The presence of a spheroidal component can be confused with a contamination of the stellar halo. Outside the Galactic centre, the stellar halo is made of a mixture of accreted~\citep[Gaia-Sausage-Enceladus merger debris][]{2018MNRAS.478..611B, 2018ApJ...863..113H, 2018Natur.563...85H} and heated disc stars~\citep[Splash/Plume in-situ populations][]{2020MNRAS.494.3880B,2019A&A...632A...4D} which, as recently demonstrated, have different kinematics and chemical abundances; hence they can be disentangled from the in-situ spheroid showing pre-disc chemical composition and the lack of rotation~\citep[Aurora, pre-disc in-situ populations][]{2022MNRAS.514..689B,2023MNRAS.525.4456B}, unless it was spun up into cylindrical rotation~\citep{1997A&A...327..983F, 2016A&A...588A..42S,2013MNRAS.430.2039S,2012MNRAS.421..333S, 2015A&A...577A...1D, 2012MNRAS.421..333S,2014A&A...567A.122D}. However, we need to underline that the discoveries regarding accreted and in-situ stellar halo components are mostly based on the samples of stars heavily weighted towards solar radius with very little knowledge/predictions on how these components should look like in the inner Galaxy, where the chemo-kinematic MW composition is governed by extremely complex boxy/peanut bulge/bar interplay. Therefore, understanding the MW bulge is crucial not only for constraining its formation and the present-day structure but also for helping us to isolate its contamination on the structure of accreted systems buried in the heart of our Galaxy.

This paper, a part of a series, aims to describe the present-day composition of the MW bulge using the orbit superposition approach and the APOGEE data. In the previous papers, we have introduced the method using mock MW data~\citep{Mapping-model}~(hereafter \citetalias{Mapping-model}) and demonstrated its advantage in the reconstruction of the survey footprint-independent chemo-kinematic structure of the MW disc~\citep{Mapping-disk}~(hereafter \citetalias{Mapping-disk}), where we also showed that the APOGEE DR 17 giants sample is already sufficient for obtaining the unbiased stellar mass-weighted MDF together with morphology and stellar kinematics across the entire MW. The paper is structured as follows. In Section~\ref{sec3::data_method} we briefly describe the methodology and the APOGEE dataset, which were deeply covered in the previous papers of the series. Section~\ref{sec3::bulge_results} focuses on the spatial structure of the bulge recovered using our orbit superposition approach. The kinematics and orbital composition of the bulge populations are presented in Section~\ref{sec3::bulge_kinematics}. Section~\ref{sec3::bulge_chemistry} covers the elemental abundance structure, gradients spatial variations, MDF and decomposition of the bulge population using Gaussian Mixture Models~(GMMs). We summarize our main findings in Section~\ref{sec3::summary}.

\section{Data and method}\label{sec3::data_method}
\subsection{APOGEE data}
In this work, we use the giant stars sample from the APOGEE DR 17, selecting stars without problematic flags and according to surface gravity. The details of the selection can be found in the second paper of the series~(\citetalias{Mapping-disk}) from which the input sample of stars is adopted. We use radial velocities, atmospheric parameters and stellar abundances~(\FeH and \MgFe) from the APOGEE DR17~\citep{2022ApJS..259...35A}, which were complemented by the proper motions from the \Gaia DR3  catalogue~\citep{2023A&A...674A...1G}. We use only stars with radial velocity uncertainties better than $2$~\kmps, distance errors $<20\%$ and proper motion errors better than $10\%$, as this is more critical to calculate stellar orbits. In order to cover a larger area across the MW disc, we select giant stars with $\rm log g < 2.2$ and flagged them as $\rm ASPCAPFLAG=0$ and $\rm EXTRATARG=0$.  

The selection of giant stars from the APOGEE catalogues, despite their stellar parameters being less precise compared to those of dwarfs, is driven by the necessity to cover a larger area of the MW~\citep[see the comparison in][]{2023ApJ...954..124I}. As demonstrated in our previous work~(\citetalias{Mapping-model}), this broader coverage is crucial for the effective application of the orbit superposition method.Here, we also do not remove stars with relatively high abundance uncertainties from our sample, as they are being used to propagate the chemical information along the orbits of the stars.

In the present analysis, we do not include stellar age information, as it may lack sufficient precision for smaller samples in the inner disc, particularly for low-metallicity and high-$\alpha$ populations. Additionally, ages do not provide significant new insights due to the well-established tight age-metallicity relation observed in high-resolution spectroscopic datasets of the bulge~\citep{2013A&A...549A.147B, 2016A&A...593A..82H, 2017A&A...605A..89B, 2018A&A...618A..78H}, hence the metallicity can be safely used to trace the structure populations of different age. In our case, the age catalogue from \cite{2024AJ....167...73S} excludes recommended ages for stars with $\FeH < -0.7$ dex, which represents about $5\%$ of the total sample and $\approx 13\%$ of the inner ($<3.5$ kpc) MW mass (see Fig.~\ref{fig03::bulge_AMR}). Although the broad age range ($\approx 6-12$ Gyr) and its distribution are consistent with the above-mentioned studies, we remain cautious about delving into the finer details of the bulge's age structure. Furthermore, we have examined several large age catalogues in our orbit superposition analysis~\citep{2019MNRAS.489.2079L, 2023MNRAS.522.4577L, 2023A&A...673A.155Q} and found quality issues affecting a substantial number of stars, limiting their utility in exploring the MW bulge.

\subsection{Orbit superposition method}

In this work, we use the output of the orbit superposition results obtained in the previous paper of the series~(\citetalias{Mapping-disk}) while the method is described in detail in \citetalias{Mapping-model}. Hence, we briefly mention the key steps of the approach. We adopt the 3D mass distribution of the MW, including its stellar component, from \cite{2022MNRAS.514L...1S}, which is an updated analytic model of the potential constructed by \cite{2017MNRAS.465.1621P}. This analytic potential, available in AGAMA~\citep{2019MNRAS.482.1525V}, shows the correct behaviour of the mass distribution outside the bar region and reproduces well the 3D density of the bar, including the X-shape structure of the bulge~\citep{2013MNRAS.435.1874W, 2015MNRAS.450.4050W}.

We integrate orbits of the APOGEE stars, assuming a constant bar pattern speed of $\rm 37~km~s^{-1}~kpc^{-1}$, in agreement with various different studies~\citep{2017MNRAS.465.1621P, 2019MNRAS.490.4740B, 2019MNRAS.488.4552S, 2020A&A...634L...8K, 2022MNRAS.512.2171C, 2022A&A...663A..38K}. The weights of the orbits in the rotating rest frame were calculated by adjusting their total 3D density to the analytic solution for the stellar component from \cite{2022MNRAS.514L...1S}. Each orbit was transformed to 500 phase-space coordinates\footnote{We increased the resolution to 2000 data points for the orbital frequency analysis in Section~\ref{sec3::bulge_kinematics}.}, and chemical abundances for these data points were assigned using normal distributions using the uncertainties as a width of the distribution. In other words, each orbit can be considered as a sample of stars following each other along the orbit with similar stellar parameters. 

We remind that our approach heavily relies on the assumption about the dynamical equilibrium of the MW disc. While there is evidence that this is not entirely true near the Sun~\citep[see, e.g.][]{2011MNRAS.412.2026S, 2013MNRAS.436..101W, 2018A&A...616A..11G, 2018Natur.561..360A} and in the outer disc of the MW~\citep[see, e.g.][]{2018MNRAS.481L..21P, 2021A&A...649A...8G, 2024arXiv240718659P}, so far, we have no clues of ongoing disequilibrium associated with the ongoing X-shaped structure evolution or the bar growth. The apparent evolution of the bar may include spiral-bar interconnection~\citep{2020MNRAS.497..933H, 2024MNRAS.528.3576V} resulting in a short time scale of the bar parameters, as they are quasi-periodic, depending on the beat frequency of spirals and the bar; however, the time-averaged properties of the bar remain unchanged. A possible ongoing deceleration of the MW bar caused by the interaction with the DM halo~\citep{1998ApJ...493L...5D,2000ApJ...543..704D} can be more problematic for our model. For instance, \cite{2021MNRAS.500.4710C} estimated the current bar slowing rate as of $\rm \approx~-4.5~km~s^{-1}~kpc^{-1}~Gyr^{-1}$, potentially resulting in $\approx 10\%$~($\approx 300-450$~pc) increase of the bar size over the last Gyr. Note, however, that for the $\approx 9$~Gyr old MW bar~\citep{2024MNRAS.530.2972S, 2024A&A...690A.147H}, such a rapid recent transformation is rather unlikely, as $N$-body models clearly show much slower bar parameters change $1-2$~Gyr after its formation~\citep{1998ApJ...493L...5D, 2000ApJ...543..704D, 2003MNRAS.341.1179A,2019A&A...629A..52L,2021A&A...650L..16F}. Finally, the deceleration rate was obtained by \cite{2021MNRAS.500.4710C} using a pure $m=2$ bar explaining the existence of the Hercules stream by the bar deceleration; the latter, however, is not needed to shape the solar neighborhood kinematics if the bar density distribution accounts also for an $m=4$ component~\citep{2018MNRAS.477.3945H}. Considering the high level of uncertainties about the time-dependence or ongoing evolution of the MW bar parameters, we can safely assume that the inner MW is in dynamic equilibrium, obviously apart from the rigid rotation of the bar itself, which we consider in the present work.

\section{Bulge structure and orbital composition}\label{sec3::bulge_results}
\subsection{3D density distribution in the inner MW}\label{sec3::bulge_structure}

In order to demonstrate the advantages of the orbit superposition approach in Fig.~\ref{fig03::bulge_den_vs_APOGEE}, we compare the obtained projected stellar density map with the footprint of the APOGEE DR 17 in the Galactic coordinates~($l,b$). The top panels show the entire data sample and total MW density, while the bottom ones are zoomed into the bulge region, within $<3.5$~kpc from the Galactic centre. Since APOGEE is a near-IR survey, it covers well the inner MW region but still does not capture the regions very close to the midplane~($|b|<2-3^\circ$); moreover, the spatial footprint is very patchy and represented by several isolated observed fields. On the contrary, in our approach, we are able to fill all these gaps by 'painting' the abundance data from APOGEE along the orbits in the bulge, providing the full 3D chemo-kinematic picture of the inner MW, effectively correcting the spatial footprint of the survey. The importance of the complete MW bulge coverage is vital as its chemo-kinematic characteristics change a lot as a function of the sky coordinates~\citep[see, e.g.][and reference in Introduction]{2016PASA...33...27D, 2018ARA&A..56..223B, 2020IAUS..353...26G}, not surprisingly, as stellar populations experience very complex dynamical evolution~\citep[see, e.g.][]{2014MNRAS.437.1284Q, 2014A&A...567A.122D, 2019A&A...628A..11D, 2017A&A...606A..47F, 2018A&A...616A.180F, 2019MNRAS.488.3324F, 2020MNRAS.494.5936F, 2019ApJ...874...67B, 2020MNRAS.498.3334D, 2023ApJ...946..118D, 2021MNRAS.503.2203C,  2023ApJ...955...38B, 2024MNRAS.527.2919A, 2024arXiv240709799B}.

The bottom-right panel of Fig.\ref{fig03::bulge_den_vs_APOGEE} illustrates the structure of the MW bulge, as reconstructed from the orbits of real stars in the APOGEE survey. This reveals the characteristic boxy and the X-shaped bulge structure, which has been previously detected through the split in the magnitude distribution of red clump stars along specific lines of sight~\citep[see, e.g.,][]{2010ApJ...724.1491M, 2010ApJ...721L..28N, 2011AJ....142...76S} and mapped in full 3D by \cite{2013MNRAS.435.1874W}. At the same time, the density contours highlight the boxy structure and the X-shaped component, which is very prominent further away from the midplane, $b>5^\circ$. Two lobes of the bulge are asymmetric in longitude relative to the Galactic Centre because of the bar orientation relative to the Sun-Galactic Centre line, making the nearby lobe more vertically extended than the farther one.

As the bar/bulge density structure varies a lot as a function of the position in the disc, in Fig.~\ref{fig03::bulge_den_2D_structure}, we present stellar column density maps in $(l, b)$ coordinates at varying distances from the Sun. From top to bottom of Fig.~\ref{fig03::bulge_den_2D_structure}, the panels display density slices that capture the asymmetry in the disc caused by the Galactic bar and bulge. The maximum density shifts from left to right as we slice the bar at different distances from the Sun. At around $-4$ kpc from the Galactic centre, the near side of the bar becomes visible as an overdense region with a moderate thickening, shifted by approximately $l \approx - 20^\circ$ relative to the Galactic centre. As we approach $-2.7$ kpc from the centre, the bulge reveals a distinctive diamond-like structure aligned with the bar, which appears most prominent at around $-1.3$ kpc from the Galactic centre. The middle panel demonstrates the symmetric, boxy component of the inner bulge. On the far side~(three bottom panels), the bar/bulge structure mirrors itself across the longitude axis, appearing less extended in the latitude direction due to geometric effects. 

In Fig.~\ref{fig03::bulge_den_1D_structure} to highlight the ability of our orbit superposition method in the reconstruction of the details of the density distribution, we present the one-dimensional stellar density distributions along various lines of sight toward the Galactic bulge. The top-left panel illustrates the positions of $5\times11$ fields ($l = -15^\circ$ to $15^\circ$) selected at five latitudes ($b = 0^\circ, 3^\circ, 7^\circ, 10^\circ, 15^\circ$), overlaid on the stellar density map. Notably, two density peaks are already visible at $b = 15^\circ$, emphasizing the vertical extent of the X-shaped bulge structure. While shifting towards each other, these peaks become more pronounced at lower latitudes and are predominantly observed for $l > 0^\circ$~(represented by shades of red lines).
The increase in separation between the peaks of density for greater heights above the plane has been used as a manifestation of the peanut structure of the MW bulge~\citep{2012ApJ...756...22N, 2014A&A...569A.103R, 2015A&A...584A..46G, 2016PASA...33...27D, 2016A&A...589A.122G} as it is shown in various MW-like barred galaxy simulations~\citep{2010ApJ...720L..72S,2017MNRAS.469.1587D, 2017MNRAS.465.1621P, 2017A&A...607L...4F, 2018ApJ...861...88B}. The picture we observe here is consistent with other observations finding a single peak density distribution near the midplane ~\citep{2005MNRAS.358.1309B, 2007MNRAS.378.1064R, 2012ApJ...750..169P, 2013A&A...552A.110G} and becoming bimodal at higher latitudes~\citep{2010ApJ...724.1491M, 2011AJ....142...76S, 2013MNRAS.435.1874W, 2013MNRAS.430..836N}.

We conclude that our method successfully recovers the full 3D density distribution of the MW bulge, which is expected given the proven strength of orbit superposition techniques in capturing fine details of bars and bulges~\citep{2012ApJ...757L...7L, 2012MNRAS.427.1429W, 2013MNRAS.435.3437W, 2020ApJ...889...39V, 2024MNRAS.534..861T}. What sets our approach apart is its use of parameters from resolved stellar populations, enabling us to explore the bulge's structure based on the chemical composition of its sub-populations. This eliminates the need for additional assumptions regarding chemo-kinematic relations~\citep{2019MNRAS.487.3776P, 2020MNRAS.496.1579Z, 2022A&A...664A.115Z}, which is in the focus of the following sections.

\subsection{Orbits of the MW bulge}
 
The presented bulge density structure provides us with a very important conclusion that the orbit superposition approach allows us to recover the details of the MW bulge where stars from the APOGEE DR 17 giants sample reproduce precisely the orbital backbone of the X-shaped structure. A detailed description of the orbital families of the X-shaped structures in simulated systems is given in \cite{2020ApJ...895...12P}, while here we focus on the dominant components we find in the reconstructed bulge of the MW. Using the standard techniques~\citep{1991A&A...252...75P, 2003MNRAS.341.1179A, 2020ApJ...895...12P}, we calculate frequencies along the Cartesian coordinates ($f_x$ and $f_z$) and cylindrical radius~($f_r$) for all orbits passing trough the inner $3.5$~kpc. In Fig.~\ref{fig03::bulge_freq_distr}, we show the in-plane frequency ratio distribution for the APOGEE stars whose orbits have masses obtained in the orbit superposition modelling. The most prominent peak, around $f_r/f_x\approx2$, highlights the orbits supporting the bar~\citep[see, e.g.][]{2014MNRAS.445.3525P,2015MNRAS.448..713P,2019MNRAS.490.2740P}, while the rest can be considered as the disc as stars on these orbits do not follow the elongated density distribution along the bar major axis~\citep[see more details in][]{2020ApJ...895...12P}. Based on the orbital frequencies, the definition of the bar yields $\approx 11\%$ of the total MW stellar mass, or $\approx 5.86\times 10^{9}$~\Msun.

Next, we decompose the bar-like orbits according to the vertical to in-plane frequency ratio, $f_z/f_x$, shown in the bottom panel of Fig.~\ref{fig03::bulge_freq_distr}. The stellar mass-weighted distribution roughly follows the ones presented in \cite{2015MNRAS.448..713P}; however, in our case, the peak at $f_z/f_x\approx2$, corresponding to the banana-like orbits~\citep{1991A&A...252...75P, 2002MNRAS.337..578P, 2002MNRAS.333..847S} is more prominent. We find that about 36\% of the bar mass is represented by orbits of this class. Unlike \cite{2015MNRAS.448..713P} who analysed only the bar orbits~($f_r/f_x\in2\pm0.1$) in a narrow range of vertical-to-in-plane orbital frequencies ($1.4<f_z/f_x < 2.1$), we present the full orbital composition of the entire orbital frequency ratio range of the bar. In Fig.~\ref{fig03::bulge_orbits} we show the face-on and side-on density structure of the orbital families in different ranges of the $f_z/f_x$ ratio. Similarly to \cite{2015MNRAS.448..713P} one can see that the X-shaped bulge component is dominated by the orbits with $1.4<f_z/f_x<1.9$~(first four columns) corresponding to the pretzel-like orbits. In this case, the X-shaped structure has very sharp edges~\citep[see also][]{2014MNRAS.445.3525P,2018A&A...612A.114P} with a total mass of $\approx 1.7\times 10^9$~\Msun. Higher order orbital members~($1.9<f_z/f_x<3$) represent the peanut component, where the banana-like~($f_z/f_x\approx 2$) orbits dominate. The highest-order orbits correspond to the edges of the bar major axis. In our case, the fraction of the X-shaped orbits is somewhat lower, $\approx 30\%$ versus $44\%$ found by \cite{2015MNRAS.448..713P}. This discrepancy comes from a higher mass of the peanut and the long bar components which was increased in \cite{2017MNRAS.465.1621P}, providing a better agreement for the bar pattern speed, and propagated to the analytic potential we adapt in our modelling. We remind that the masses of the bulge components correspond to the orbital-based definition of the bar~($f_r/f_x\in2\pm0.1$) and do not include the non-resonant disc component or higher-order periodic orbits embedded into the inner Galaxy, which in fact are far less significant building blocks of bars~\citep{2019MNRAS.490.2740P}. 

\section{Bulge line-of-sight kinematics}\label{sec3::bulge_kinematics}
As discussed above, the intricate 3D density distribution and orbital structure of the MW bulge naturally lead to complex kinematic behaviour among its stellar populations. The known kinematic trends in the MW bulge region include a cylindrical rotation with the decrease and flattening of the velocity dispersions with vertical distance from the disc plane~\citep{2009ApJ...702L.153H, 2013MNRAS.432.2092N, 2014A&A...562A..66Z}. However, the details of the stellar kinematics vary with the metallicity of stellar populations~\citep{2017A&A...606A..47F, 2017MNRAS.469.1587D, 2019A&A...628A..11D, 2019MNRAS.485.5073D,2024arXiv240709799B} allowing to test various scenarios of the MW bulge formation. 

\begin{figure}[h!]
    \centering
    \includegraphics[width=1\hsize]{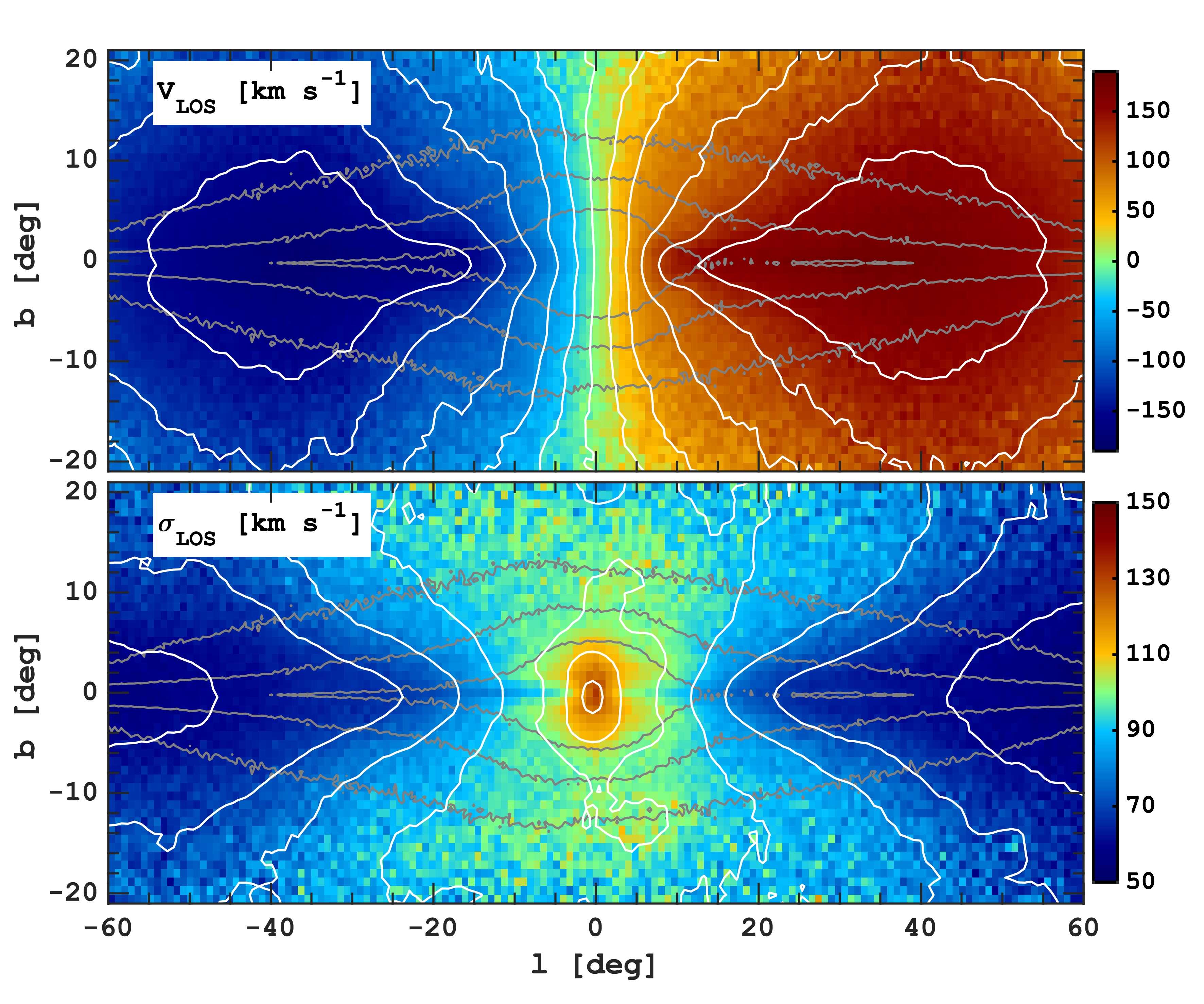}
    \caption{Kinematics of the MW bulge recovered using orbit superposition approach. The panels show the mean line-of-sight velocity~(top) and the line-of-sight velocity dispersion~(bottom) in the MW bulge region, limited by a heliocentric distance of $8.12 \pm 3.5$ kpc, around the Galactic centre. The grey contours reflect the shape of the stellar density distribution, while the white ones correspond to the constant velocity and velocity dispersion values. The bulge region shows a prominent rotation due to its disc origin, while the asymmetry of the radial velocity profile is due to projection effects. The velocity dispersion map shows a prominent peak extended perpendicular to the midplane.}
    \label{fig03::bulge_vlos2D}
\end{figure}

\begin{figure*}
    \centering
    \includegraphics[width=1\hsize]{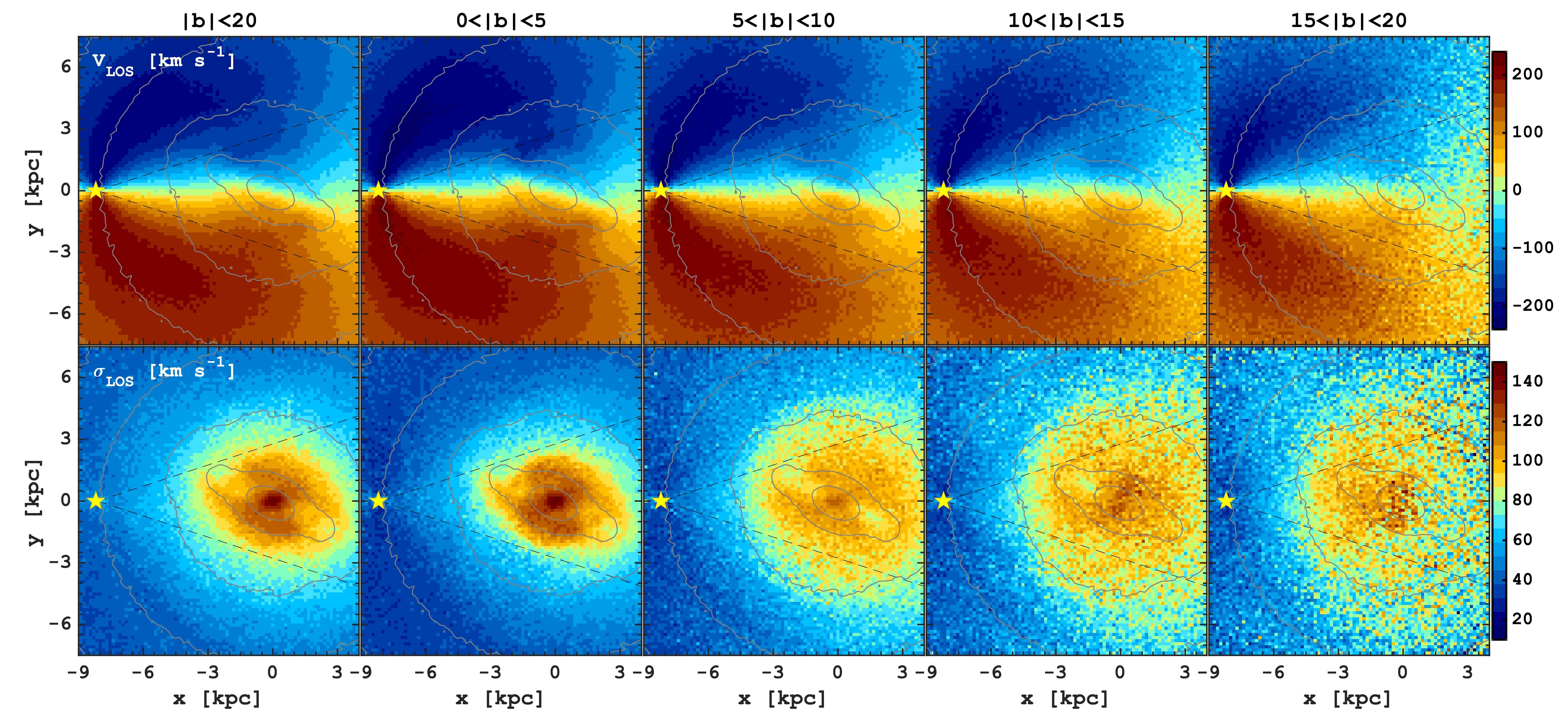}
    \caption{Kinematics of the inner MW and bulge in the face-on projection. The panels show the mean heliocentric line-of-sight velocity~(top) and velocity dispersion~(bottom) maps at different latitudes. The grey contours show the total stellar isodensity levels, marking the extent and orientation of the bar. The dashed lines highlight the longitude values of $l\pm 20^\circ$. The yellow asterisks mark the position of the Sun. }
    \label{fig03::bulge_vlos_faceon}
\end{figure*}

Before exploring the metallicity-dependent variations in bulge kinematics, we first present the overall trends. In Fig.~\ref{fig03::bulge_vlos2D}, we display the projected, averaged line-of-sight velocity and velocity dispersion for stars within a heliocentric distance of $8.12 \pm 3.5$kpc, covering the entire bulge-bar region. The kinematic maps reveal strong rotational motion, with the strength of rotation decreasing at higher latitudes. At any given latitude, populations located further out in the disc have higher line-of-sight velocities. The kinematic disc component is clearly present, as emphasized by the narrowing of iso-velocity contours toward the mid-plane. In the innermost region~($l<20^\circ$), the velocity isocontours gradually broaden towards higher latitudes. The velocity dispersion map highlights the presence of colder disc populations near the mid-plane, showing a trend with latitude. This trend is in the sense that the velocity dispersion decreases with increasing latitude. The most striking feature in the bulge region is the peak velocity dispersion, reaching up to $\approx 150$~\kmps near the Galactic centre, rapidly decreasing and saturating around $100$~\kmps at latitudes $|b|>5^\circ$. The picture we see here is in agreement with data from the BRAVA, ARGOS and APOGEE surveys, which have shown that the bulge is cylindrically rotating~\citep{2008ApJ...688.1060H, 2013MNRAS.432.2092N, 2016ApJ...819....2N}, which is the typical velocity field produced by bars. In this case, the velocity dispersion peak in the very centre does not imply the presence of the classical bulge, as it can be explained by a pure buckled bar models~\citep{2010ApJ...720L..72S, 2011MNRAS.416L..60B, 2012ApJ...757L...7L, 2013MNRAS.430.2039S, 2015A&A...577A...1D}.

While maps in the ($l,b$) perspective provide us with the projected kinematics typically available in observations, the structure of the bulge as a function of distance can help us understand the 3D motions of stars in the inner MW. Figure~\ref{fig03::bulge_vlos_faceon} shows the mean line-of-sight velocity and velocity dispersion in several latitude bins~($0-5$, $5-10$, $10-15$ and $15-20^\circ$). Such information is hard to obtain in the existing observational datasets, as it requires a large-scale spatial coverage, but more importantly, precise distance and radial velocity determination far away from the Sun. In the case of our approach, these quantities are known from the orbits of APOGEE stars and do not blur the observed picture.  The kinematic maps presented in the figure show, first of all, large-scale disc rotation, which in the centre is affected by the presence of the bar. Since the bar is inclined relative to the Sun - Galactic centre line, we can see a prominent tilt of the radial velocity associated with the streaming motion along its major axis, often observed in extragalactic IFU data~\citep{2015A&A...576A.102A, 2016A&A...591A...7G, 2016MNRAS.456..692M, 2019MNRAS.482.1733G, 2024arXiv241005374K} and simulations~\citep{2015ApJ...812L..16D,2019MNRAS.485.5073D}. As we explored in \citetalias{Mapping-disk}, such a velocity tilt is difficult to obtain using the raw Gaia or APOGEE data, as the precision of the distance determination is not sufficient in this region~\citep{2023A&A...674A..37G}. The radial velocity tilt in the bar region is the most prominent at latitudes close to the midplane, and the pattern fades away with increasing distance from the midplane.

In Fig.~\ref{fig03::bulge_vlos_faceon} the line-of-sight velocity dispersion face-on maps also reveal several peculiarities in the inner bulge region. The velocity dispersion has a peak near the Galactic centre~($<0.5$~kpc), but it disappears rapidly at latitudes higher than $b\approx 5^\circ$. Such a feature is easy to explain as this region is dominated by the shortest pretzel-like orbits, which have very complex shapes~(see Fig.~\ref{fig03::bulge_orbits}). Once stars with different orbital phases are mixed, they result in a very high line-of-sight velocity dispersion. Interestingly, the velocity dispersion decreases along the bar major axis with local minima on the edges; the latter can be understood as there is not much diversity in the orbital composition of the bar's edge, which is dominated by orbits with the highest $f_z/f_x$ ratio~(see Fig.~\ref{fig03::bulge_orbits}). The radial velocity dispersion plateau, observed along the minor bar axis, corresponds to the X-shaped component.

\begin{figure*}
    \centering
    \includegraphics[width=1\hsize]{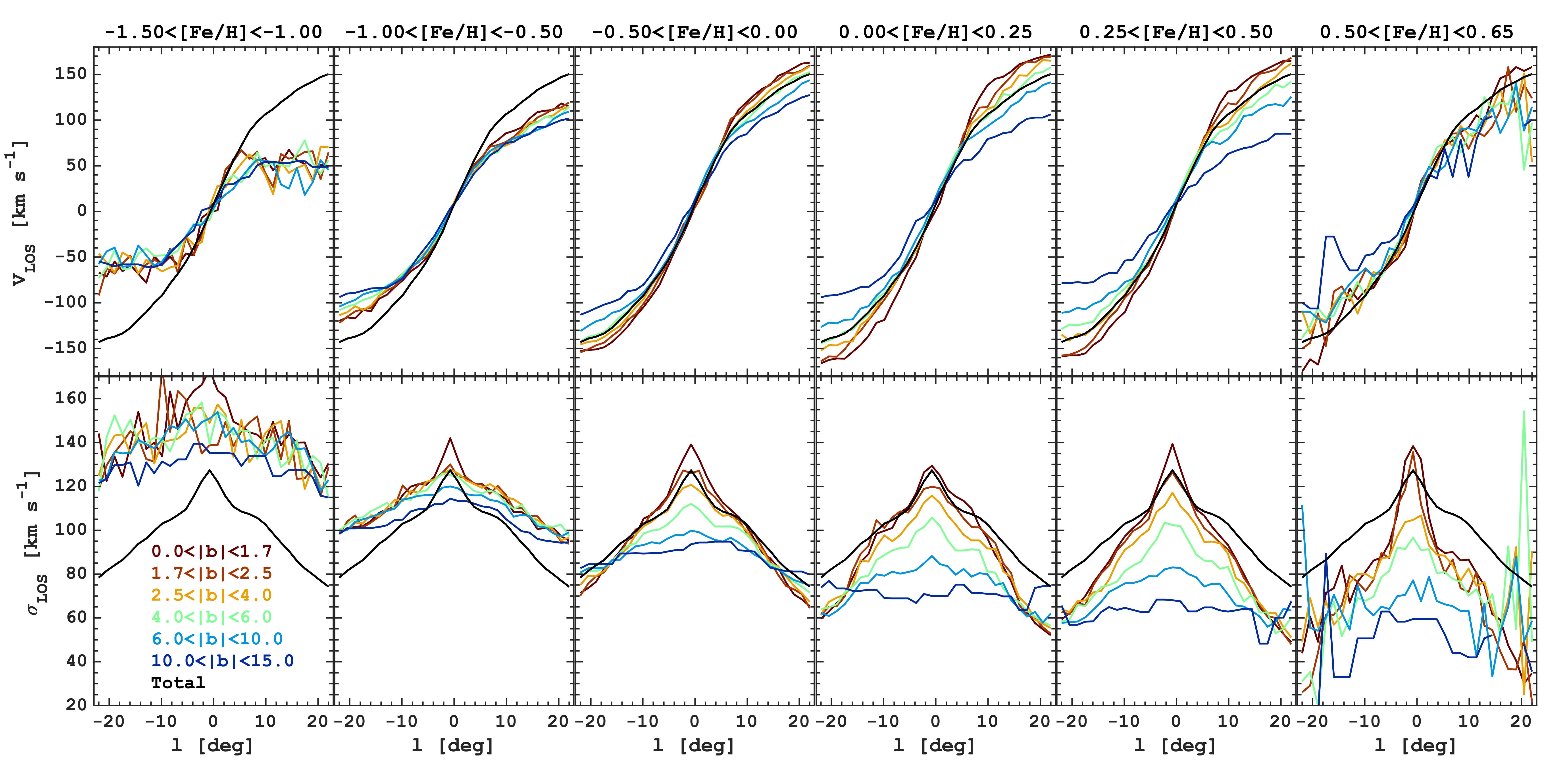}
    \caption{Line-of-sight kinematics of the MW bulge. The panels show the mean line-of-sight radial velocity~(top) and the line-of-sight velocity dispersion~(bottom) versus longitude measured at different latitudes and in \FeH bins, as marked in the header of the top panels. The line colour indicates the latitude range. The panels are based on the data within $< 3.5$~kpc around the Galactic centre. In each panel, the black lines show the averaged velocity and velocity dispersion profiles for all stellar populations of the entire region.
     }
    \label{fig03::bulge_los1D}
\end{figure*}

Next, we proceed with a more conventional analysis of the MW bulge kinematics that is widely used in the literature. In Fig.~\ref{fig03::bulge_los1D} we show the line-of-sight velocity and velocity dispersion longitude profiles for different latitude slices~(different colour) and also separated for populations of different metallicity. For reference, in each panel, the total velocity and velocity dispersion profiles for all stars located inside the bulge~($<3.5$~kpc) are shown with black lines. Although the MW bulge MDF is known to possess several components, the metallicity bins were chosen arbitrarily for the current analysis, while the complete MDF decomposition is investigated in Section~\ref{sec3::bulge_chemistry_populations}. The following trends can be observed in Fig.~\ref{fig03::bulge_los1D}:
\begin{itemize}

    \item In all the panels, the rotation is cylindrical within $|l|<10^\circ$, with very little difference in rotation speed at different latitudes, except for the highest latitude~($b=10-15^\circ$, dark blue lines) showing a shallower velocity gradient in this longitude range. This behaviour is observed in all metallicity bins, except for the most metal-poor one $\FeH<-1$, showing almost negligible difference in rotation as a function of latitude. Such uniform cylindrical rotation is observed in many models where the bulge is produced via vertical instability of the bar~\citep{2010ApJ...720L..72S, 2017A&A...606A..47F, 2019MNRAS.485.5073D}. We note that almost no rotation is found for the very metal-poor bulge stars~\citep[below $\FeH \approx -1.5$; ][]{2020MNRAS.491L..11A}, which the APOGEE sample does not account for; however, we estimate the total mass of this population is less than $1\%$.

    \item For a given metallicity range~(except $\FeH<-1$), the velocity profiles diverge at larger longitudes, $l>10$, showing the decrease of the velocity~(or shallower velocity gradient) with increasing latitude, as expected due to asymmetric drift. The velocity profiles are nearly symmetric around $l=0$, with a velocity being slightly higher for $l>0$ compared to $l<0$~\citep[see also][]{2007A&A...465..825C,2009ApJ...691.1387R,2013A&A...555A..91V,2014A&A...562A..66Z}. The latter is likely the result of the projection effect where the far side of the bulge is seen at lower latitudes~(see Fig.~\ref{fig03::bulge_den_2D_structure} and Fig.~\ref{fig03::bulge_vlos_faceon}). A weak longitudinal asymmetry of the velocity dispersion can also be seen in the bottom panels. 

    \item There is a trend in increasing velocity outside $l \simeq 10^\circ$ with increasing metallicity from $\FeH=-1.5$ to $\approx 0$~dex, which then is weakly reversed, as seen for the higher latitudes. In a scenario where the low and high-metallicity bulge populations correspond to the thick and thin discs, respectively, this is explained if thin disc stars lose more angular momentum once trapped by the boxy/peanut bulge than thick disc populations~\citep{2017A&A...606A..47F}~(see more details and comprehensive analysis in \citealt{2024arXiv240709799B}).

    \item The velocity dispersion gradually decreases with increasing metallicity, as one can expect, assuming a reasonable age-velocity dispersion relation and increase in metallicity with age. This, however, does not mean that the observed velocity dispersion remained unaffected by the bulge emergence, as suggested by \cite{2017MNRAS.469.1587D} the differential impact of the vertical bar instability on pre-existing disc populations results in their kinematic fractionation, preserving but not completely erasing the ``initial'' kinematics~\citep[see also][]{2017A&A...606A..47F,2019A&A...628A..11D}. 

    \item For a given metallicity range, at high latitudes~($b>5-6^\circ$) the velocity dispersion profiles are almost flat. At latitudes $<5$ we can see that the velocity dispersion increases towards $l=0^\circ$; however, thanks to the panoramic coverage of the reconstructed MW bulge data, inaccessible for the sparse raw spectroscopic datasets, we can see a prominent flattening of the profile around $l \approx 10^\circ$, seen in simulations of MW-like bulge galaxies~\citep{2017MNRAS.469.1587D}. This flattening of the radial velocity dispersion profile reflects the plateau we discussed in Fig.~\ref{fig03::bulge_vlos_faceon}, which we attribute to the X-shaped component. 
    
    \item The central velocity dispersion peak, as discussed above, is caused by the inner boxy bulge component. Interestingly, its peak value slightly increases from solar towards very metal-rich populations, suggesting that they are trapped more efficiently in the boxy/peanut instability, which can be the case if they were initially colder compared to less metal-rich populations~\citep{2017MNRAS.469.1587D,2017A&A...606A..47F}. These findings provide an interpretation of the recently re-discovered metal-rich knot in the innermost MW~\citep{2024arXiv240601706R}.

\end{itemize}

\begin{figure*}[h!]
    \centering
    \includegraphics[width=1\hsize]{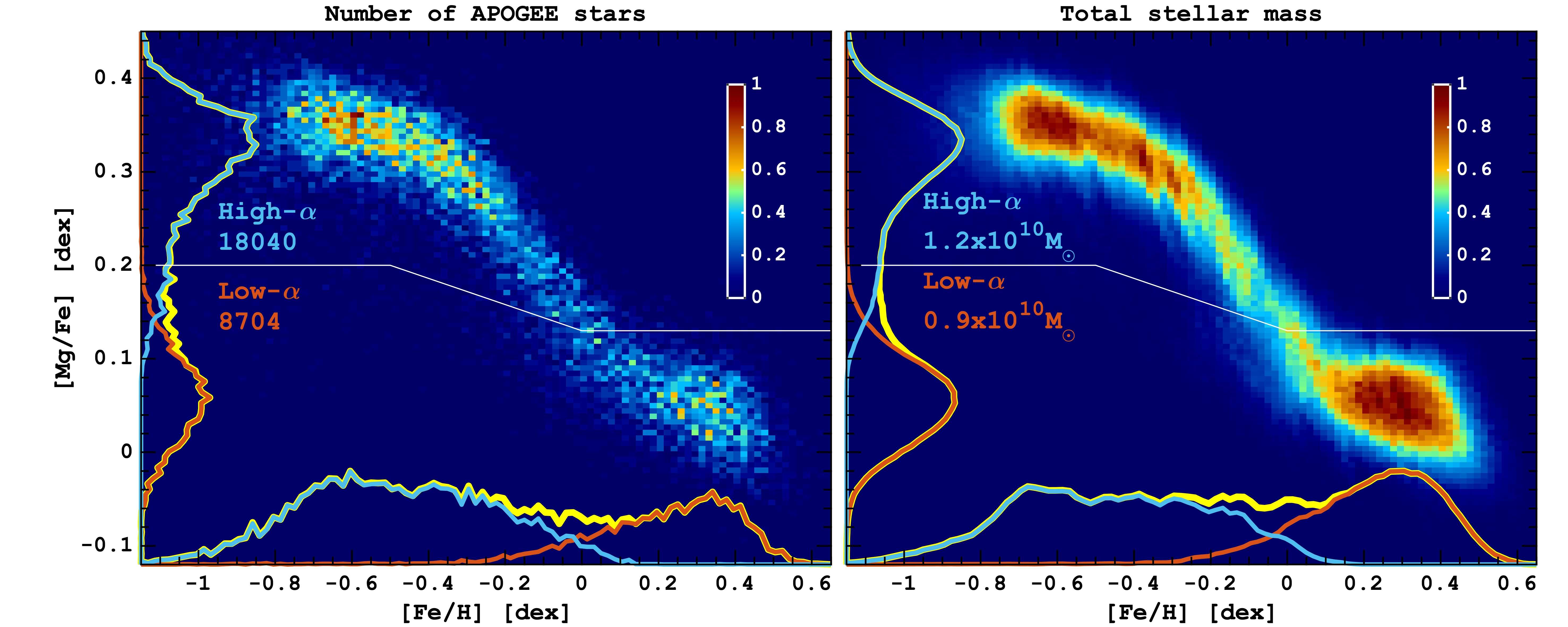}
    \caption{\MgFe-\FeH relation for the MW bulge. The left panel shows the distribution of APOGEE stars, while the right panel depicts the mass-weighted distribution obtained using the orbit superposition method within $<3.5$~kpc from the Galactic centre. The yellow lines show the distribution functions of \MgFe and \FeH separately. The solid white line shows the border used to separate high- and low-$\alpha$ populations. The number of stars and stellar mass corresponding to the high- and low-$\alpha$ populations are marked in the panels. While the orbit superposition method assigns more mass to the metal-rich populations, both APOGEE data and the orbit superposition-based density reveal rather similar abundance distributions.}
    \label{fig03::bulge_bimodality}
\end{figure*}


\section{Chemical abundance trends}\label{sec3::bulge_chemistry}

Various studies suggest that the MW bulge does not consist of a single component but is a superposition of multiple components~(see Introduction) or perhaps a continuum of stellar populations with different chemical and kinematic properties shaped by both varying physical conditions of the corresponding star formation and redistribution during dynamical evolution of the bar and bulge. In this section, using the stellar mass distribution obtained in the orbit superposition reconstruction, we dissect the MW bulge in terms of its chemical abundance profiles, gradients, and distribution functions.

\subsubsection{Bulge MDF}\label{sec3::bulge_chemistry_mdf}

In Figure \ref{fig03::bulge_bimodality}, we compare the raw APOGEE data with the reconstructed stellar mass-weighted distribution in the \FeH-\MgFe plane for the region within $<3.5$~kpc. The white line divides the distribution into high- and low-$\alpha$ populations, following the method used in the analysis of the MW disc~(\citetalias{Mapping-disk}). The relative mass distributions of the high- and low-$\alpha$ populations along the \FeH and \MgFe axes are represented by the blue and red lines, respectively, while the total distribution functions for \MgFe and \FeH are shown in yellow. The MW bulge closely resembles the chemical abundance pattern of the MW disc, which is expected, as the bulge is composed of thin~(mostly low-$\alpha$) and thick~(mostly high-$\alpha$) disc stars, as discussed earlier~\citep[see, e.g.][]{2010A&A...512A..41B, 2011A&A...530A..54G, 2014AJ....148...67J, 2014A&A...562A..71B}.

The fractional contribution of the high- and low-$\alpha$ components in the bulge region is reversed compared to what is observed across the entire MW disc. In particular, around 40\% of the bulge consists of low-$\alpha$ populations, whereas approximately 44\% of the total MW mass is high-$\alpha$, based on the same definition~(\citetalias{Mapping-disk}). It is important to note that this reversal is not due to differences in radial density profiles, as the mono-abundance populations in the bulge have similar scalelengths of $\approx 2$ kpc~(\citetalias{Mapping-disk}). Additionally, the orbit-superposition-based mass fraction for the bulge shows a higher proportion of high-$\alpha$ stars compared to the ratio derived from the raw APOGEE data, where about $32\%$ is low-$\alpha$. This discrepancy arises because our approach allows us to account for regions of the bulge near the midplane, which are underrepresented in the APOGEE footprint (see Fig.~\ref{fig03::bulge_den_vs_APOGEE}). Therefore, we conclude that the \MgFe distribution is bimodal with relatively close fractions of low- and high-$\alpha$ populations~\citep[see also][]{2019A&A...626A..16R}.

\begin{figure}[h!]
    \centering
    \includegraphics[width=1\hsize]{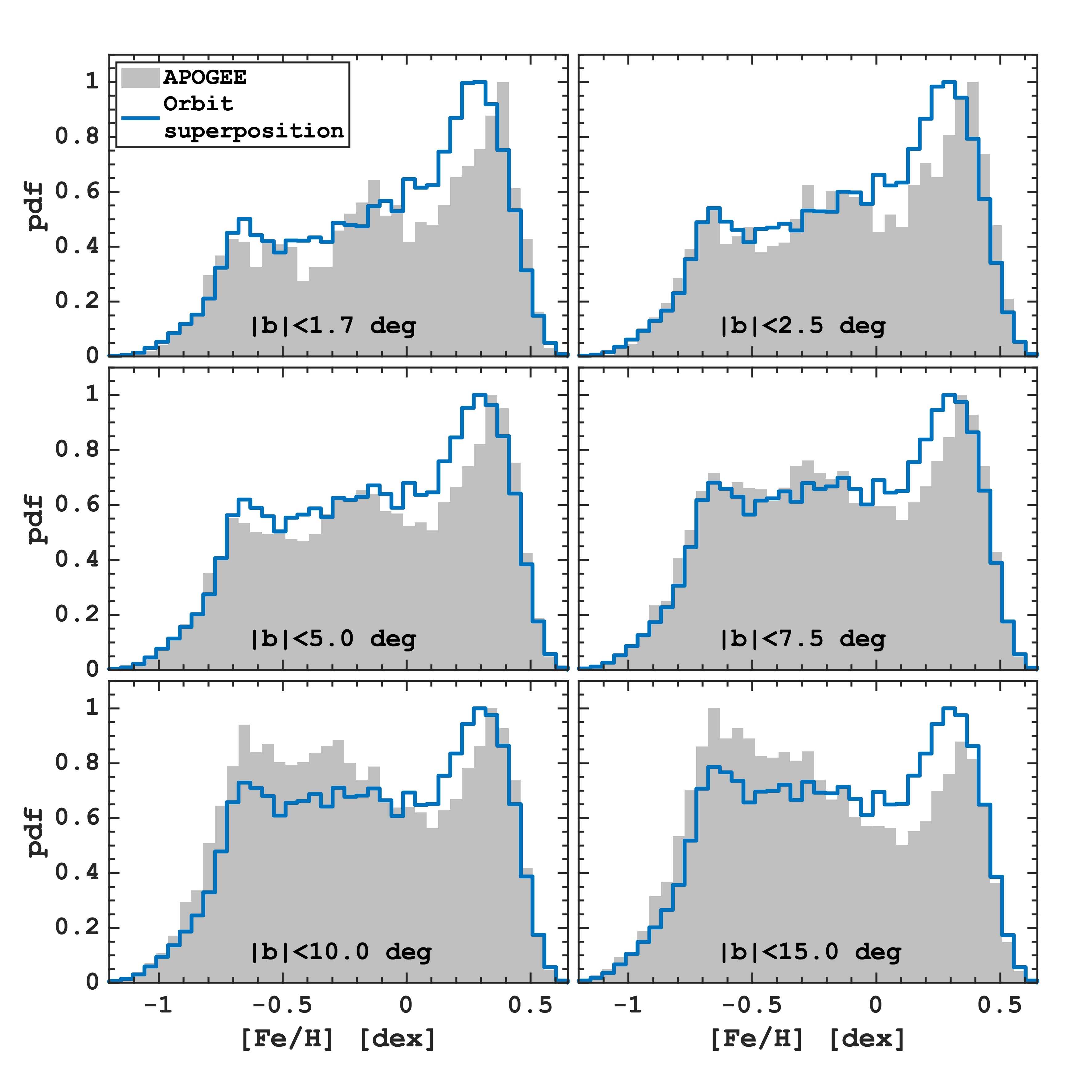}
    \caption{Comparison of the MW bulge MDF limited by different distances in the vertical direction from the midplane as marked in each panel. The blue line is the MDF weighed using the orbit superposition method versus the raw APOGEE sample counts in grey. The orbit superposition method and APOGEE show similar MDFs close to the midplane; however, some systematic differences become prominent once the input sample covers regions further out.}
    \label{fig03::bulge_MDF_APOGEE}
\end{figure}

Although the stellar mass-weighed MDF~(right panel in Fig.~\ref{fig03::bulge_bimodality}) is very wide, it is characterised by two main components, centred at $-0.7$ and $+0.3$~dex, corresponding to the centres of distributions of the high- and low-$\alpha$ disc populations. Several weak modulations are observed between these metallicity values, potentially indicating the presence of several components claimed by \cite{2013MNRAS.430..836N, 2010A&A...512A..41B, 2011A&A...533A.134B, 2013A&A...549A.147B, 2017A&A...605A..89B}. Before delving into the details of the bulge MDF, we provide more comparisons with the raw APOGEE data. In Fig.~\ref{fig03::bulge_MDF_APOGEE}, we show the MDF of the bulge stars inside $R<3.5$~kpc, limiting our spatial selections by different latitude values. 

Overall, we notice that the global shape of the MDFs for the raw APOGEE and stellar mass-weighted datasets very closely resemble each other and can be broadly described as bimodal~\citep{2011A&A...534A..80H, 2012A&A...546A..57U, 2011A&A...533A.134B, 2015A&A...584A..46G,2015A&A...583L...5G, 2017A&A...599A..12Z, 2014A&A...569A.103R, 2017A&A...601A.140R, 2017A&A...600A..14S}. However, there are several small but significant differences. In all the panels, the metal-rich peak is shifted to slightly lower values in the orbit superposition results than the raw APOGEE. Near the midplane ($b<1.7-5^\circ$), the APOGEE MDF shows 3-4 peaks~\citep{2013MNRAS.430..836N}, while the mass-weighted method fills the gaps between them. The best match we observe is for the middle right panel ($|b|<7.5^\circ$). Once we consider populations at higher latitudes, the APOGEE MDF lacks a number of metal-poor stars, which is seen in the bottom panels. We conclude that the correction of the MDF provided by the orbit superposition is rather modest, highlighting a very good APOGEE selection function in the bulge region, weakly affecting the metallicity distribution relative to underlying mass distribution~\citep{2017A&A...601A.140R, 2017AJ....154..198Z, 2017A&A...606A..97N}. 

Using the ARGOS spectroscopic survey of multiple fields in the bulge region, \cite{2013MNRAS.430..836N}  identified five components in the bulge MDF and associated them with different stellar populations, with decreasing metallicity: A – a thin, boxy/peanut-shaped~($\FeH \approx +0.15$); B – a thicker, boxy, peanut-shaped~($\FeH \approx -0.25$); C – the inner thick disc~($\FeH \approx -0.7$); D – the metal-weak thick disc~($\FeH \approx -1.2$); and E – the inner stellar halo~($\FeH \approx -1.7$). The five bulge MDF components were identified by \cite{2013A&A...549A.147B} using microlensed dwarfs with a good agreement with the ones from \cite{2013MNRAS.430..836N}~\citep[see also][for the most recent results and comparison]{2017A&A...605A..89B}. The agreement of these studies, based on different datasets with different spatial coverage, supports the existence of multiple distinct bulge populations with different spatial distributions and kinematics. In the case of the orbit superposition, while we notice some small amplitude modulation between the main high- and low-$\alpha$ components, it is not obvious whether these MDF features can be associated with the components known in the literature.  

\begin{figure}[h!]
    \centering
    \includegraphics[width=1\hsize]{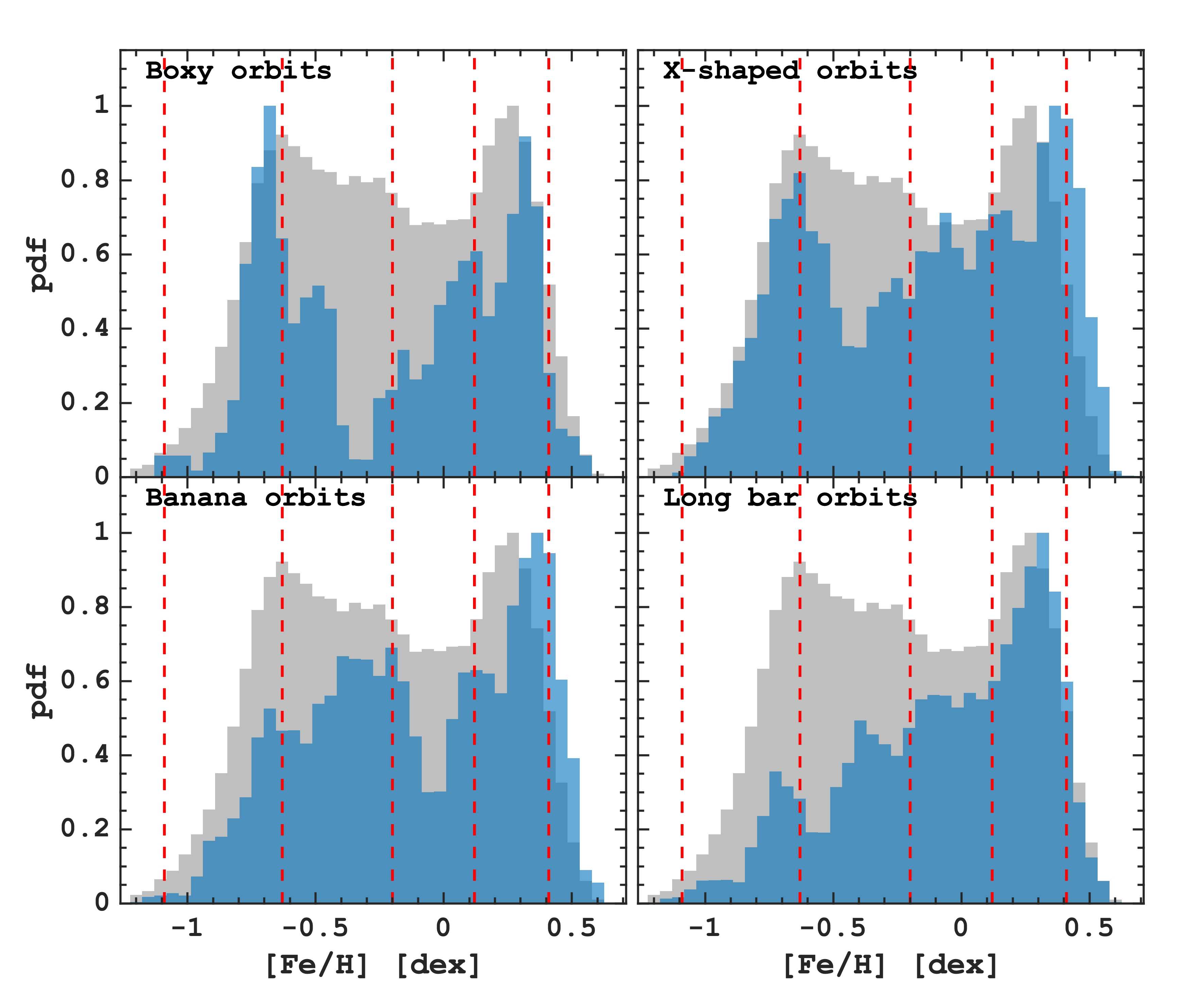}
    \caption{MDFs of different orbital families of the MW bulge. The orbital families are classified according to the in-plane to vertical frequencies ratio: inner boxy orbits~($f_z/f_x = 1.4-1.6$), X-shaped orbits~($f_z/f_x=1.6-1.9$), banana orbits~($f_z/f_x=1.9-2.2$) and long bar orbits~($f_z/f_x>2.2$) orbits. In each panel, the distributions are normalized to the maximum value.  For reference, the vertical red dashed lines mark the locations of the peaks in the MDF based on the microlensed dwarf and subgiant stars in the MW bulge from \cite{2017A&A...605A..89B}. The same grey histogram in the background shows the MDF of the disc orbits with $|f_z/f_r-2|>0.1$.}
    \label{fig03::bulge_MDF_orbits}
\end{figure}

Since we have identified stars in the APOGEE dataset whose orbits maintain the MW bulge boxy/peanut structure, it is natural to associate corresponding orbital families with the MDF features.  In different panels of Fig.~\ref{fig03::bulge_MDF_orbits} we present the MDF of the orbits in several orbital frequency ratio ranges~(see Fig.~\ref{fig03::bulge_orbits}), representing the inner boxy orbits~($f_z/f_x = 1.4-1.6$), X-shaped orbits~($f_z/f_x=1.6-1.9$), banana orbits~($f_z/f_x=1.9-2.2$) and long bar orbits~($f_z/f_x>2.2$). The red vertical lines highlight the location of the MDF components from \cite{2017A&A...605A..89B}. The grey histogram shows the MDF of the disc stars, i.e. stars with orbital frequencies which do not support the bar, $|f_z/f_r-2|>0.1$.

We find that all orbital families have broad metallicity distribution with multiple peaks whose location differs from one orbital family to another. Some of the peaks match the position of components from \cite{2017A&A...605A..89B}. We need to keep in mind that the metallicity scale of the APOGEE can be different from the one in \cite{2013MNRAS.430..836N}~(ARGOS) or \cite{2017A&A...605A..89B} samples~\citep{2017A&A...601A.140R}. Nevertheless, we discuss the features of individual orbital families and the differences between them. 

The inner boxy orbits are the only ones whose MDF matches reasonably well almost all peaks from \cite{2017A&A...605A..89B}, and it is the only orbital family which shows a small peak near $\FeH \approx -1$. \cite{2013MNRAS.430..836N} associated this peak~(D) with the thick metal-weak disc present in the MW before the bulge formation. The next metallicity peak, around $\FeH\approx -0.7$, is seen for all orbital families supporting the bulge. However, its significance decreases from inner boxy to X-shaped orbits, but it is still somewhat present in the banana orbits and has a very small contribution in the long bar orbital families. This MDF peak is the most dominant one, corresponding to the centre of the high-$\alpha$ populations, or inner thick disc~(component C in \citealt{2013MNRAS.430..836N}). However, it is also seen for stars not supporting the bar orbits~(see grey histogram). The two following MDF peaks~($\approx -0.2$ and $\approx +0.1$~dex) are also seen among all bulge-supporting orbital families. The most metal-rich MDF peak matches the best by the X-shaped and banana orbits; interestingly, these orbits capture more metal-rich stars compared to the background disc whose metal-rich MDF peak is visibly shifted towards lower metallicities by $\approx 0.15$~dex~\citep[see, e.g.][]{2018ApJ...861...88B}. 

To summarize, we demonstrated that the MDFs of different orbital families supporting the MW bulge indeed reveal multiple peaks, some of which can be associated with the MDF components found in the literature. On the contrary, the background disc orbits MDF is dominated by the two main rather broad components associated with high- and low-$\alpha$ populations of the bulge.

\begin{figure*}[h!]
    \centering
    \includegraphics[width=1\hsize]{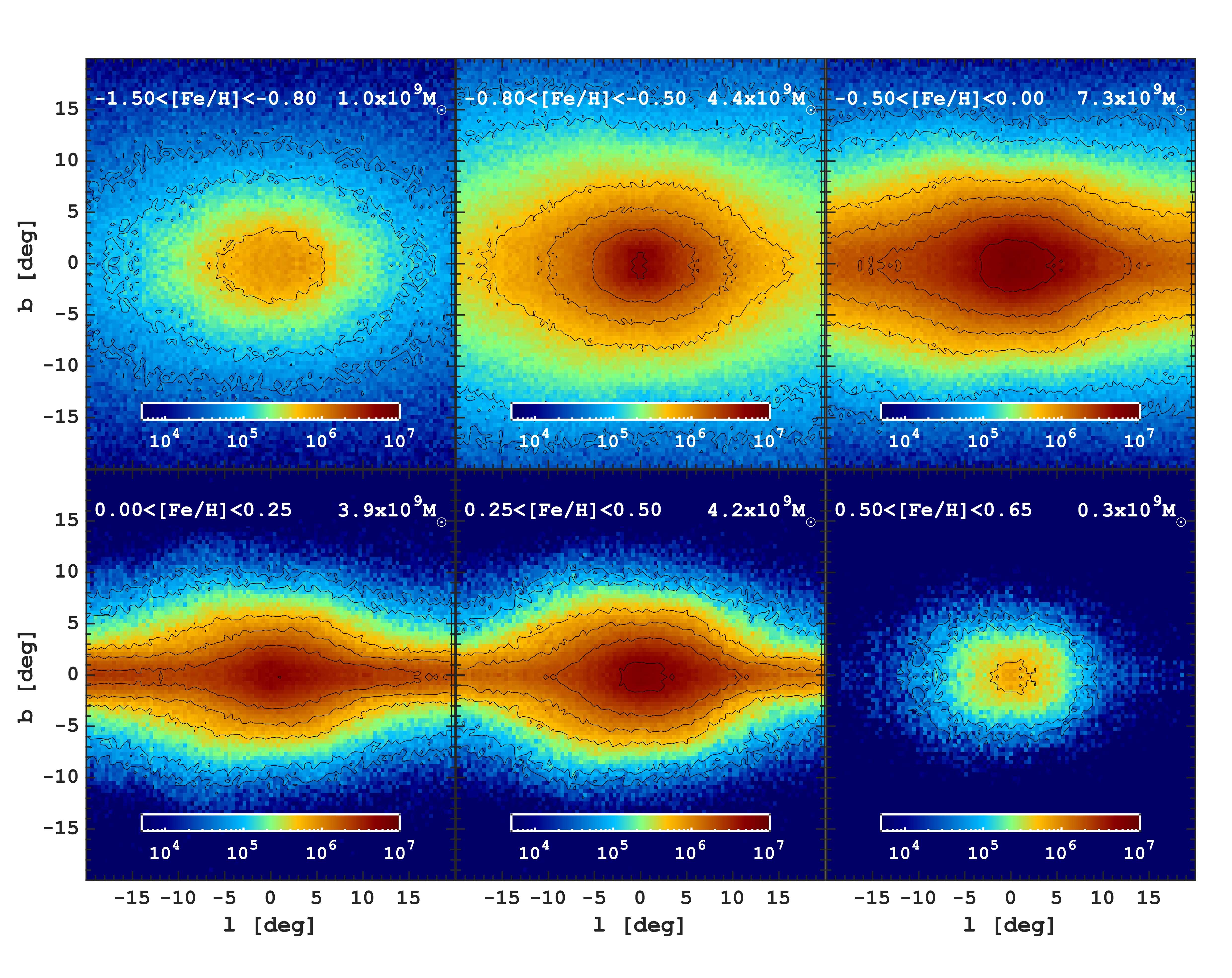}
    \caption{Shape of the MW bulge stellar density across different metallicity bins. The corresponding metallicity range and stellar mass are specified at the top of each panel. The top three panels represent the high-$\alpha$ populations, while the bottom panels depict the low-$\alpha$ populations. The mapping of these populations into the bulge reveals distinct trends where the contrast of the X-shaped/boxy structure increases with metallicity. High-$\alpha$ stars, corresponding to a kinematically hot thick disc component, are more widely dispersed in the bulge. In contrast, low-$\alpha$ stars are part of a colder population that is more closely confined to the midplane but still efficiently mapped into the bulge. The most (extremely) metal-rich populations exhibit a vertically extended, highly concentrated distribution towards the centre of the MW without displaying a prominent disc-like morphology.     
    \href{https://www.dropbox.com/scl/fi/z5k8rdocdilqwnvx412lt/evolution_bulge_density_feh_evolution.mp4?rlkey=zvbg9b12sf3r29r11cxml8t78&dl=0}{Animation}}
    \label{fig03::bulge_density_feh}
\end{figure*}

By construction, the 3D density distribution of the MW bulge is defined by the adopted potential and the precision of its recovery using the orbit superposition. However, our approach does not control the dependence of the bulge density structure (and kinematics) for populations with different abundances. Therefore, it is very interesting to see how different stellar populations are mapped into the bulge. We conclude this section by presenting the MW bulge density structure as a function of metallicity. In Figure~\ref{fig03::bulge_density_feh} we show the projected stellar density distribution in $(l,b)$ coordinates for six metallicity bins for stars inside $3.5$~kpc from the Galactic centre. For consistency, we use the same metallicity bins for which we analysed the bulge kinematics~(see Fig.~\ref{fig03::bulge_los1D}).

\begin{figure*}[h!]
    \centering
    \includegraphics[width=1\hsize]{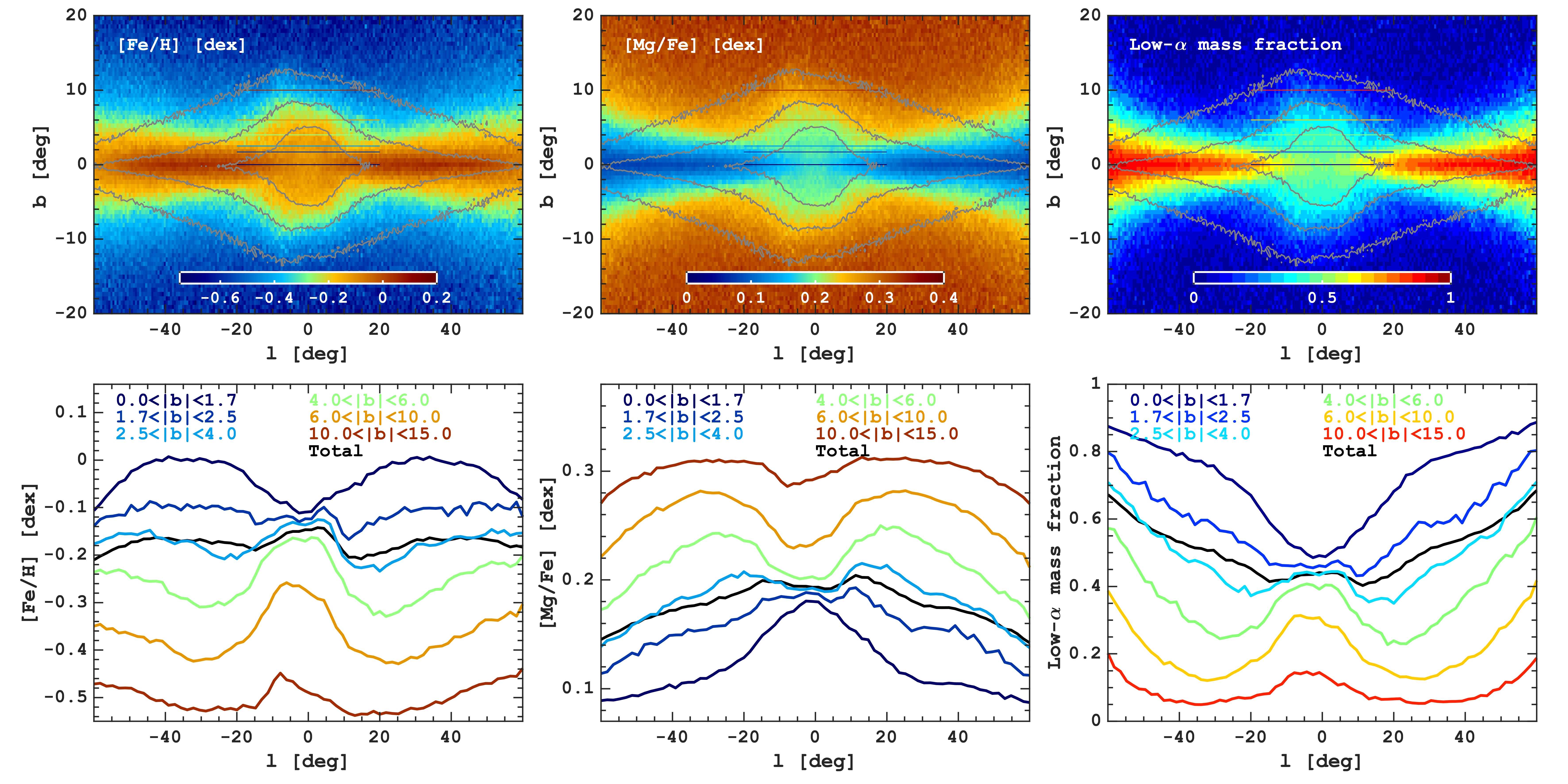}\\
    \includegraphics[width=1\hsize]{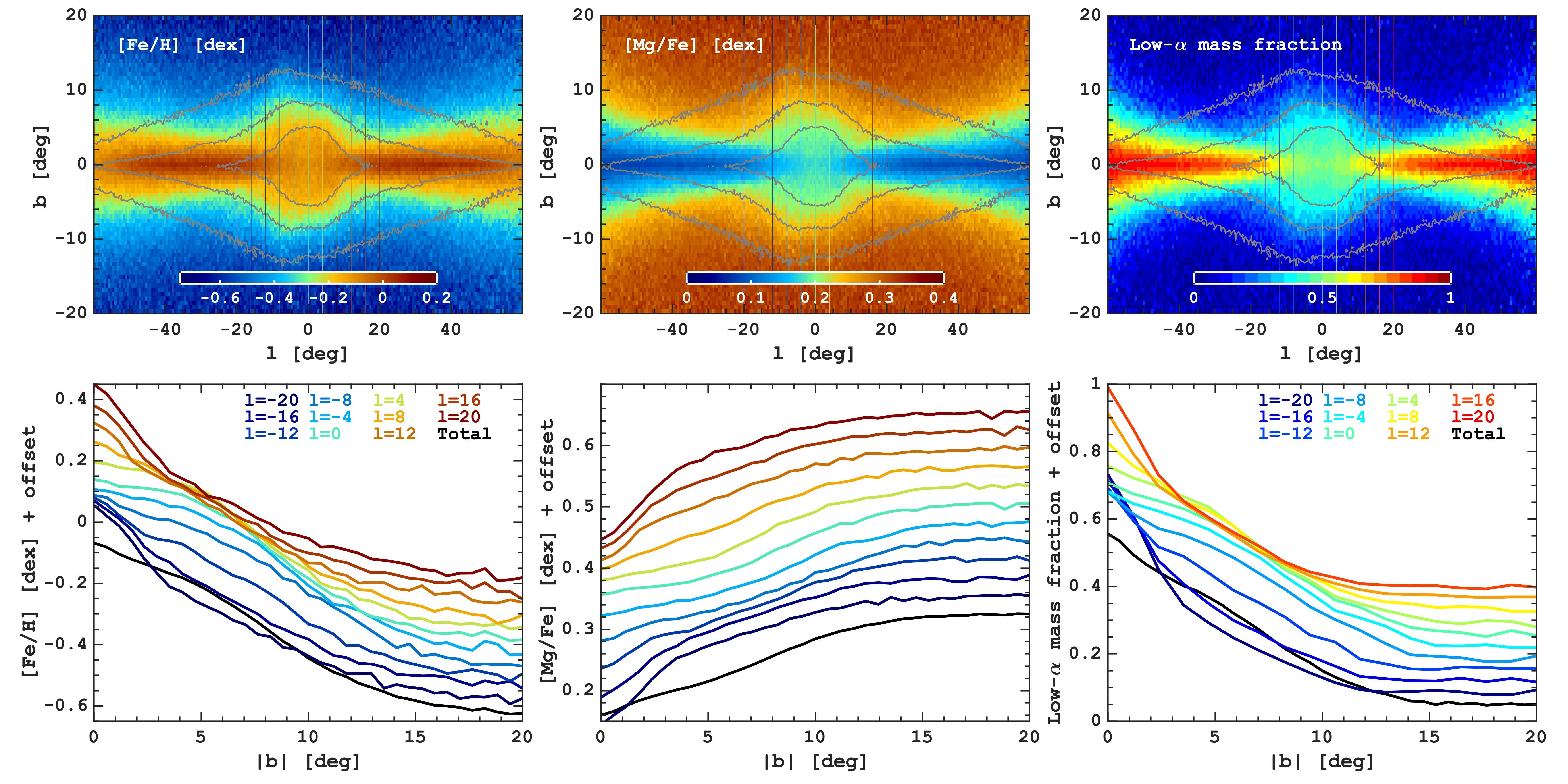}
    \caption{Chemical abundance profiles across the bulge region. The left and middle panels in the top row show the mean stellar mass-weighted \FeH and \MgFe, respectively, while the right one depicts the stellar mass fraction of the low-$\alpha$ populations~(see Fig.~\ref{fig03::bulge_bimodality} for the definition). The bottom panels show one-dimensional profiles of the same quantities as a function of longitude and at different latitudes, marked with different colours. The bottom group of panels shows the same, but for the vertical profiles, along latitude. The black lines at the bottom show the latitude- or longitude-averaged profiles. The grey background lines in the 2D abundance maps depict the stellar isodensity contours.}
    \label{fig03::bulge_los_chem}
\end{figure*}

\begin{itemize}
    \item The most metal-poor bin~($-1.5 < \FeH <-0.8$) shows an oval-like distribution elongated along the midplane but with no clear signature of the discy or the X-shaped component. Since these populations are the most metal-poor stars in our APOGEE sample, and they are likely contaminated by the halo~(in-situ and accreted) stars~\citep{2019A&A...632A...4D,2020MNRAS.494.3880B} whose ages are typically $>10-11$~Gyr. Moreover, their initial hot orbits make it hard for them to be captured by the X-shaped structure, which should appear not earlier than the MW bar itself, which is $\approx 8-9$~Gyr old~\citep{2024MNRAS.530.2972S,2024A&A...690A.147H}. Nevertheless, as they show little rotation~(see Fig.~\ref{fig03::bulge_los1D}), they are still dominated by the metal-poor, early MW disc potentially corresponding to the spin-up phase~\citep{2022MNRAS.514..689B, 2024ApJ...962...84S, 2024arXiv240816815Z}. 

    \item The following two metallicity bins ($-0.8 < \FeH <-0.5$ and $-0.5 < \FeH <-0$) correspond to the bulk of the high-$\alpha$ sequence or the inner thick disc. This is reflected by their large vertical extension. Both populations demonstrate the boxy component and the X-shaped structure is more prominent for the more metal-rich bin.

    \item Two supersolar metallicity bins~(metal-rich $0<\FeH<0.25$ and very metal-rich $0.25<\FeH<0.5$) represent the bulk of the low-$\alpha$ stars and, as expected, the initially thin disc populations, are more tightly confined to the midplane. However, towards the Galactic centre, the thickness increases sharply, making rise to the bulge, where the X-shaped structure is the most prominent among all populations. Interestingly, the very metal-rich populations appear to be thicker than the metal-rich ones. This difference could be attributed to the vertical bar instability, which  impacts the colder (very metal-rich) populations a bit more efficiently, pushing them further from the midplane. In contrast, the metal-rich stars were already kinematically hotter at the time of bulge formation, making them less susceptible to this effect. 

    \item The last, extremely metal-rich~ populations~(EMR, $\FeH>0.5$) reveal a very compact, boxy structure without a strong signature of the X-shaped component. These stars show a very thin and compact disc component outside the bulge region~\citep{2024arXiv240601706R} and, while being the most metal-rich population of the MW formed about $5-9$~Gyr ago~(\citetalias{Mapping-disk}), can be formed only in a disc-like configuration with little radial redistribution over time. The appearance of these compact populations as a vertically thick component simply reflects their influence by the vertical bar instability, even if they formed after the emergence of the main X-shaped/boxy component~\citep{2019MNRAS.485.5073D}. Therefore, this apparent spheroid-like shape of the EMR stellar populations distribution do not suggest the precents of the classical bulge.

 
\end{itemize}


\begin{figure}[h!]
    \centering
    \includegraphics[width=1\hsize]{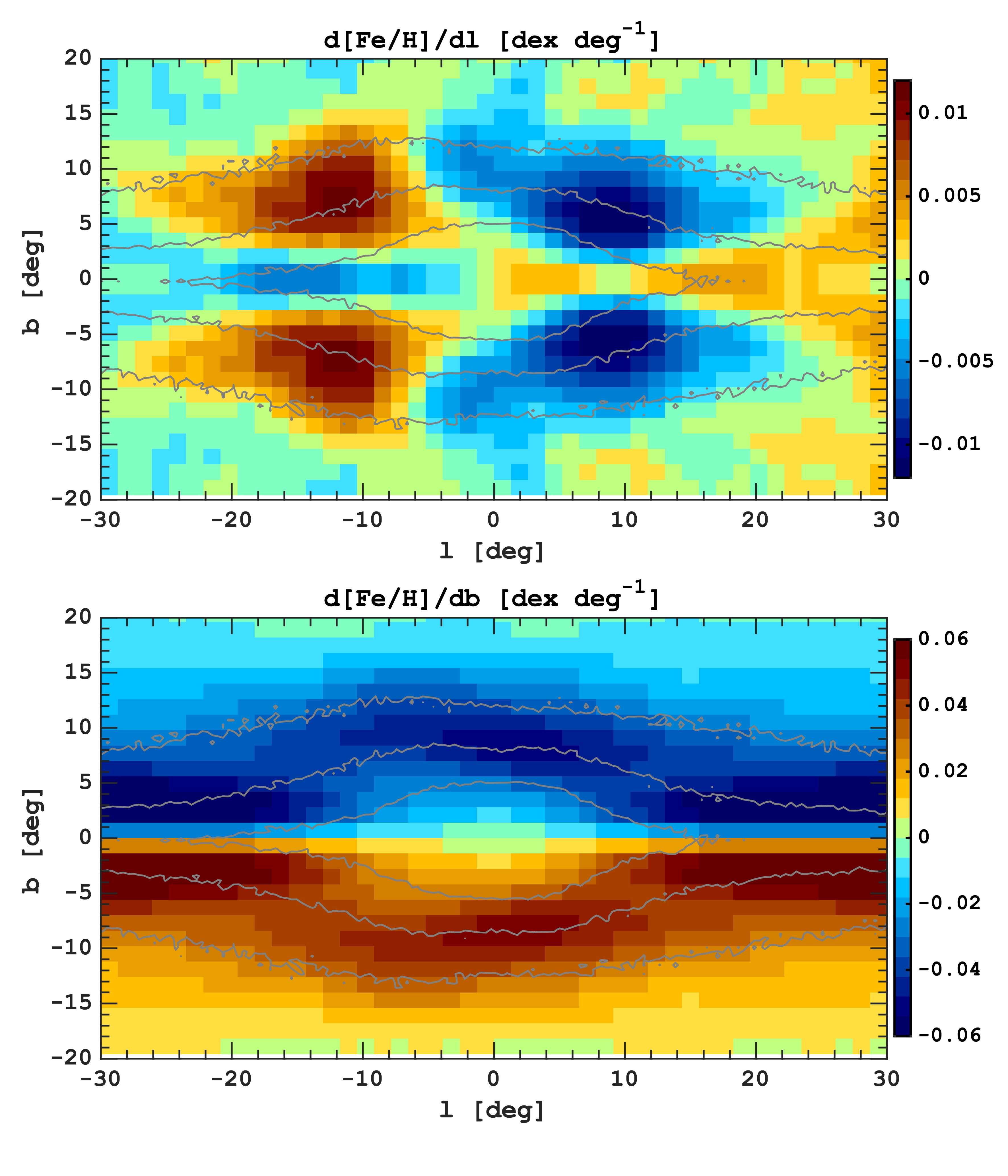}
    \caption{Maps of the radial~(top) and vertical~(bottom) metallicity gradients in the MW bulge region. The maps show the slopes of the metallicity profiles measured in each pixel within $\pm 1$~deg in the corresponding direction. The radial and vertical gradient variations trace well the X-shaped and boxy density components, respectively. The gradients are steeper near the transition regions between high- and low-$\alpha$ populations. The grey background lines reflect the projected stellar density distribution.}
    \label{fig03::bulge_los_gradients}
\end{figure}

\subsection{Abundance variations and gradients in bulge}\label{sec3::bulge_chemistry_gradients}

The determination of the abundance gradients and spatial variations in the bulge is fundamental because its exact shape can be related to several different bulge formation channels.
The presence of the vertical metallicity gradient within the inner $2$ kpc was demonstrated by \cite{1995MNRAS.277.1293M}. Later, \cite{2007ApJ...665L.119R} found no evidence of major iron abundance or abundance ratio gradient between the inner field and Baade window, suggesting the lack of the vertical metallicity gradient. On the contrary, \cite{2008A&A...486..177Z} demonstrated the existence of a vertical metallicity gradient along the bulge minor axis and the photometric metallicity map of \cite{2013A&A...552A.110G} indicates a strong vertical gradient of $\rm \approx -0.28~dex~kpc^{-1}$. The secular MW-like bulge formation models suggest that the metallicity gradients are inevitable because of the pre-existing radial and vertical metallicity gradients in the MW disc(s) stellar populations, which are then modified, depending on the kinematics of corresponding populations, by the X-shaped structure but can not be erased~\citep{2017MNRAS.469.1587D}. 

The comparison of the exact values of the bulge metallicity gradient is challenging as different surveys cover different regions of the bulge with varying depths along the line of sight. With the orbit superposition data, we are no longer affected by the survey spatial footprint and can analyse the abundance variations across the whole bulge region. In Fig.~\ref{fig03::bulge_los_chem}, we present the maps of the mean \FeH, \MgFe and low-$\alpha$ mass fraction. The latter is useful for understanding the contribution of the main bulge~(disc) components in different regions. The mean metallicity and \MgFe maps show very complex distributions, which, in fact, follow the density distribution~(grey background contours). The metallicity is higher in the densest regions close to the midplane and decreases away from it, implying a negative metallicity gradient. The central region of the bulge seems to have a flat metallicity distribution with the mean metallicity around $\approx -0.1$, which is somewhat lower compared to the regions outside the bulge~($|l|>20^\circ$).  This might make the impression that the bulge is more metal-poor compared to the surrounding disc populations, or there is a positive radial metallicity gradient in the inner MW. This behaviour has been demonstrated in several spectroscopic studies demonstrating that the MW bulge has a significant metal-poor component at all latitudes~($|l, b| < 10^\circ$), which leads to the Milky Way bulge being on average metal-poor~\citep{2013MNRAS.430..836N, 2016PASA...33...22N, 2017A&A...601A.140R,2017A&A...599A..12Z}. We argue, however,  that this is a projection effect because, although the radial metallicity profile is flattened in the inner $3-4$~kpc, the radial gradient is still negative along the bar major axis~(see Fig.~21 in \citetalias{Mapping-disk}). This is supported by the Auriga simulations~\citep{2020MNRAS.494.5936F} where boxy/peanut bulges are predominantly metal-rich, in agreement with IFU spectroscopic studies of the inner regions of external galaxies~\citep{2017MNRAS.466L..93G, 2020A&A...637A..56N} and, hence, the MW is a typical barred galaxy.

At higher latitudes~($|b|>2.5^\circ$), the metallicity variations along longitude are not monotonic~(see Fig.~\ref{fig03::bulge_los_chem}). The metallicity drops once we move inwards~(along longitude) from the disc to the bulge region, and it rises again once we cross the bulge. This is the result of the vertical thickening of the metal-rich low-$\alpha$ populations whose fraction is about $50\%$ at these latitudes in the bulge. Notably, the low-$\alpha$ stars are also expected to thicken but at a higher rate compared to the colder metal-rich populations, as predicted in the secular bulge formation scenario. On top of the non-monotonic metallicity variations, we observe an asymmetric behaviour relative to $l=0$. This is caused by the bar orientation where the nearby lobe is seen as a more vertically-extended metallicity distribution at $l<0$ compared to the outer lobe~($l>0$). 

\begin{figure*}[h!]
    \centering
    \includegraphics[width=1\hsize]{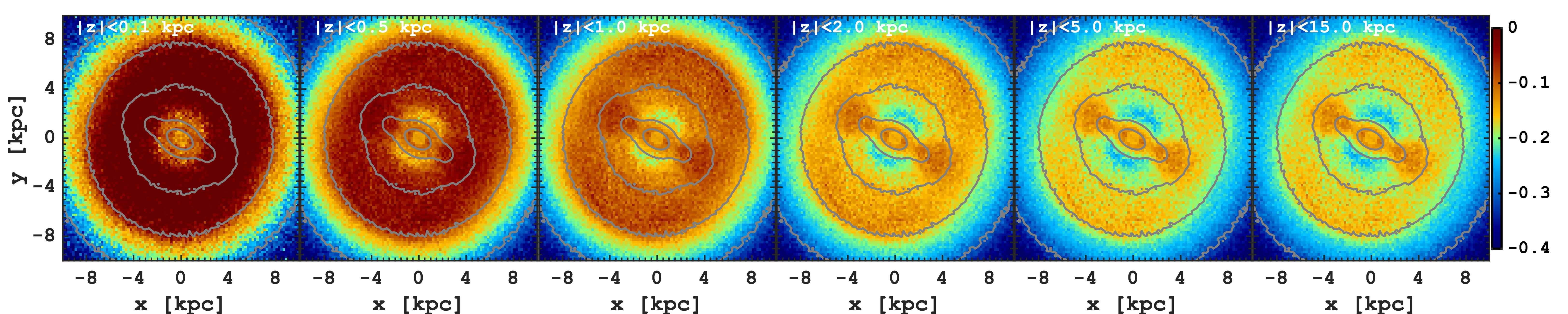}
    \caption{Face-on maps of the mean stellar metallicity \FeH distribution across the MW disc. From left to right, the maps represent stellar populations within horizontal slabs at increasing distances from the midplane. Near the midplane~(two panels on left), the bulge region appears to be metal-poor because of the deficit of the metal-rich~(low-$\alpha$) populations expelled to higher distances perpendicular to the disc plane via bar/bulge instability. However, the actual metallicity of the MW bulge is higher compared to the surrounding disc once its full vertical extension is taken into account. The bar appears to be metal-rich along the major axis because kinematically colder low-$\alpha$~(metal-rich) populations are more efficiently captured by the bar's gravitational potential compared to the hotter high-$\alpha$~(metal-poor) stars dominating along the bar minor axis.}
    \label{fig03::bulge_metallicity}
\end{figure*}

In Figure~\ref{fig03::bulge_los_chem} the \MgFe ratio variations show the inverted pattern compared to the \FeH profiles, which is a trivial result coming from the \MgFe-\FeH relation where metal-poor stars have higher \MgFe values and vice-versa~(see Fig.~\ref{fig03::bulge_bimodality}). 

To assess the metallicity gradient variations across the bulge in Fig.~\ref{fig03::bulge_los_gradients} we show the 2D maps of the radial~(along longitude) and vertical (along latitude) slopes of the metallicity profiles measured in each pixel within $\pm 1$~deg. One can see that the vertical or longitudinal metallicity gradients in the MW bulge can not be characterised by a single value. Thus, it is unsurprising that surveys covering various parts of the bulge arrive at different solutions. The vertical gradient is flat in the very centre ($l<5$ and $|b|<2$). Further out, we observe the increase of the vertical negative metallicity gradient value with both latitude and longitude. The maximum values of the vertical gradient are observed near the edge of the X-shaped structure, between the regions dominated by low- and high-$\alpha$ populations~(see rightmost panels in Fig.~\ref{fig03::bulge_los_chem}). The longitudinal metallicity gradients also change with the position, tracing the X-shaped bulge component well.

The existence of the metallicity gradients in the bulge is predicted by $N$-body and hydrodynamical simulations of secular bulge formation~\citep{2013ApJ...766L...3M,2017A&A...607L...4F,2020MNRAS.494.5936F}. The question is whether this is a pure effect of the mapping in the vertical direction of horizontal~\citep{2011ApJ...734L..20M} or vertical ~\citep{2011ApJ...738....4B,2017A&A...607L...4F} metallicity gradients initially present in the disc, or a combination of the two~\citep{2015A&A...577A...1D,2019A&A...628A..11D}. Answering this question will require reproducing the abundance maps presented in this section in simulations. However, we can not exclude certain degeneracy between different scenarios, especially if both thin and thick MW discs existed and both had radial and vertical gradients before the X-shaped structure formation. The question of how young stars populate the boxy peanut structure after its emergence is poorly explored in simulations~\citep[but see][]{2019MNRAS.485.5073D}, but it is exciting and may help to provide additional constraints on the origin of the MW bulge and conditions in the MW at the time of its emergence.

\subsection{MW bulge: metal-rich or metal-poor}\label{sec3::bulge_metal_rich}

In this section, we address a simple but rather debatable issue in the literature: whether the MW bulge is metal-poor or metal-rich. In this context, the bulge metallicity should be considered relative to the surrounding disc populations. If it is metal-poor, the MW would be pretty unusual among the nearby barred galaxies with pseudo-bulges~\citep{2020A&A...637A..56N, 2024MNRAS.534.2438N} and contrary to what is suggested in modern galaxy-formation simulations~\citep{2019ApJ...874...67B, 2020MNRAS.498.3334D, 2020MNRAS.494.5936F}, suggesting that centres of bars are metal-rich. The metal-poor MW bulge would also suggest that the radial metallicity distribution of the MW disc is not typical and has a broken profile, as reported in \citep{2023NatAs...7..951L,2019MNRAS.490.4740B}. However, in the previous work of the series, we demonstrated that this is not entirely true once the selection function and masses of mono-abundance populations are taken into account~(\citetalias{Mapping-disk}). We found that the radial metallicity gradient of the MW, although flattened in the inner region, is still negative, and the MW bar appears to be metal-rich. 

\begin{figure*}[h!]
    \centering
    \includegraphics[width=1\hsize]{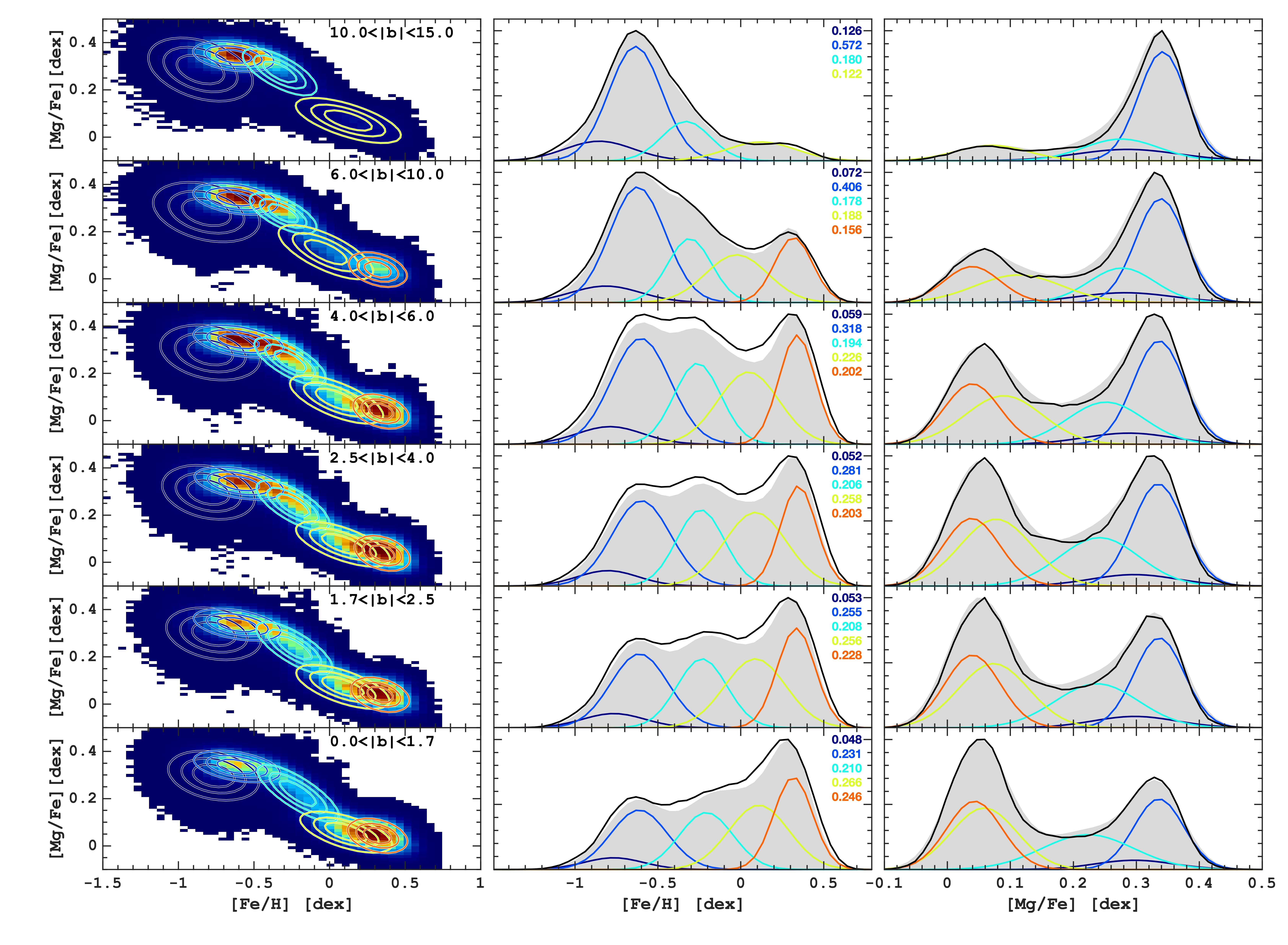}
    \caption{Two-dimensional decomposition of the MW bulge in the \MgFe-\FeH plane as a function of latitude based on the data inside $3.5$~kpc from the Galactic centre. The left panels show the normalized stellar density maps, where the contours of different colours show the components identified using the 2D GMM approach. The grey histograms in the middle and right columns show the generalized \FeH- and \MgFe- distribution functions, respectively. The coloured lines show the one-dimensional projections of the 2D-GMM components, while the black lines represent the sum of these components. In the middle row, the fractional contribution of each component is marked by the numbers of the same colour. Our analysis suggests that the \MgFe-\FeH bulge composition is best represented by the five GMM components, where only two are the main ones and correspond to the bulk of high- and low-$\alpha$ populations. Two intermediate components allow the decomposition of the transition between the main ones by fitting the $\alpha$-knee and likely do not correspond to distinct bulge populations. The most metal-poor GMM component likely corresponds to the accreted populations.}
    \label{fig03::bulge_gaussians2D}
\end{figure*}

The MW bulge metallicity structure is hard to analyse in full 3D, as the existing spectroscopic surveys have limited coverage in the inner MW; however, it is not an issue for the orbit superposition reconstructed MW. In Fig.~\ref{fig03::bulge_metallicity}, we show the mean stellar metallicity in the face-on disc projection~($XY$), considering different distances from the midplane. In other words, the metallicity is averaged in horizontal slabs with increasing~(from left to right) thickness from $0.1$~kpc to a given value indicated in each panel. Close to the midplane~($|z|<0.1-0.5$~kpc, two leftmost panels), the mean metallicity is low near the Galactic centre compared to the surrounding disc region, as expected. However, when we consider populations farther away from the midplane, an inverted trend emerges, revealing the bulge as a metal-rich component. This behaviour can be explained by assuming that the bulge consists of both a thin, metal-rich population and a thicker, metal-poor one. The deficit of metal-rich stars near the midplane in the bulge region results from vertical bar instability, which causes these populations to be ``puffed up'' above the plane. In contrast, the more metal-poor stars, which are kinematically hotter, remain less affected by this instability during bulge formation and dominate at smaller distances from the midplane while experiencing less vertical thickening. As we move away from the plane, we capture the metal-rich stars that were displaced from the midplane, increasing the average metallicity in the bulge region compared to the surrounding disc. This inversion in the metallicity profiles is seen along a longitude in Fig.~\ref{fig03::bulge_los_chem}. The explanation we present here clarifies the findings of \cite{2021A&A...656A.156Q}, who claimed that the MW bulge is metal-poor~(see their Fig. 7). This conclusion is naturally influenced by the fact that their APOGEE sample is biased toward the midplane and does not account for the APOGEE footprint or selection function properly. As a result, \cite{2021A&A...656A.156Q} obtained a metallicity distribution similar to the leftmost panels in Fig.~\ref{fig03::bulge_metallicity}, which, as we now understand, does not present a complete picture.

While the MW bulge has slightly subsolar metallicity overall, it still appears as a metal-rich component, similar to (pseudo) bulges observed in external galaxies~\citep{2006MNRAS.371..583M, 2014A&A...570A...6S, 2018A&A...614A..48B, 2024MNRAS.534.2438N}. However, in certain projections -- such as in Galactic coordinates ($l, b$) shown in Fig.~\ref{fig03::bulge_los_chem} -- the bulge can appear metal-poor due to these projection effects and also because it is being compared to the metal-rich thin disc stars in the midplane. In Fig.~\ref{fig03::bulge_metallicity} we also see that kinematically colder, metal-rich stars are more concentrated along the bar’s major axis, producing the metal-rich bar. The latter, however, does not imply the metal-rich stars formed along the bar major axis as the star formation is usually suppressed in the bar region~\citep{2018A&A...609A..60K, 2019A&A...628A..24G, 2021MNRAS.507.4389G, 2020MNRAS.499.1116F}. The latter is seen as the lack of recent significant star formation in the inner~($<4-5$~kpc, but excluding the central molecular zone) MW as well~\cite[see, e.g.][]{2022ApJ...941..162E, Mapping-disk}. Instead, the ability of younger, more metal-rich populations to be trapped by the bar more efficiently compared to the older metal-poor counterparts results in the more metal-rich MW bar, as suggested by the kinematic fractionation mechanism~\citep{2017MNRAS.469.1587D, 2017A&A...606A..47F, 2018A&A...611L...2K} and in agreement with extragalactic observations~\citep{2024MNRAS.534.2438N}.

\subsection{Chemo-kinematics of bulge populations: 2D GMM decomposition}\label{sec3::bulge_chemistry_populations}
In this section, we investigate the chemo-kinematic composition of the MW bulge populations, which are defined using the GMMs of the MDF as widely done in the literature. In our case, thanks to the complete coverage of the bulge and the knowledge of the masses of different mono-abundance populations instead of using the MDF only, we use the GMM approach to define the bulge populations in the two-dimensional \MgFe-\FeH plane. This allows us to link the chemically defined populations of the bulge with the conventional chemical abundance structure of the MW disc(s).

\begin{figure*}
    \centering
    \includegraphics[width=1\hsize]{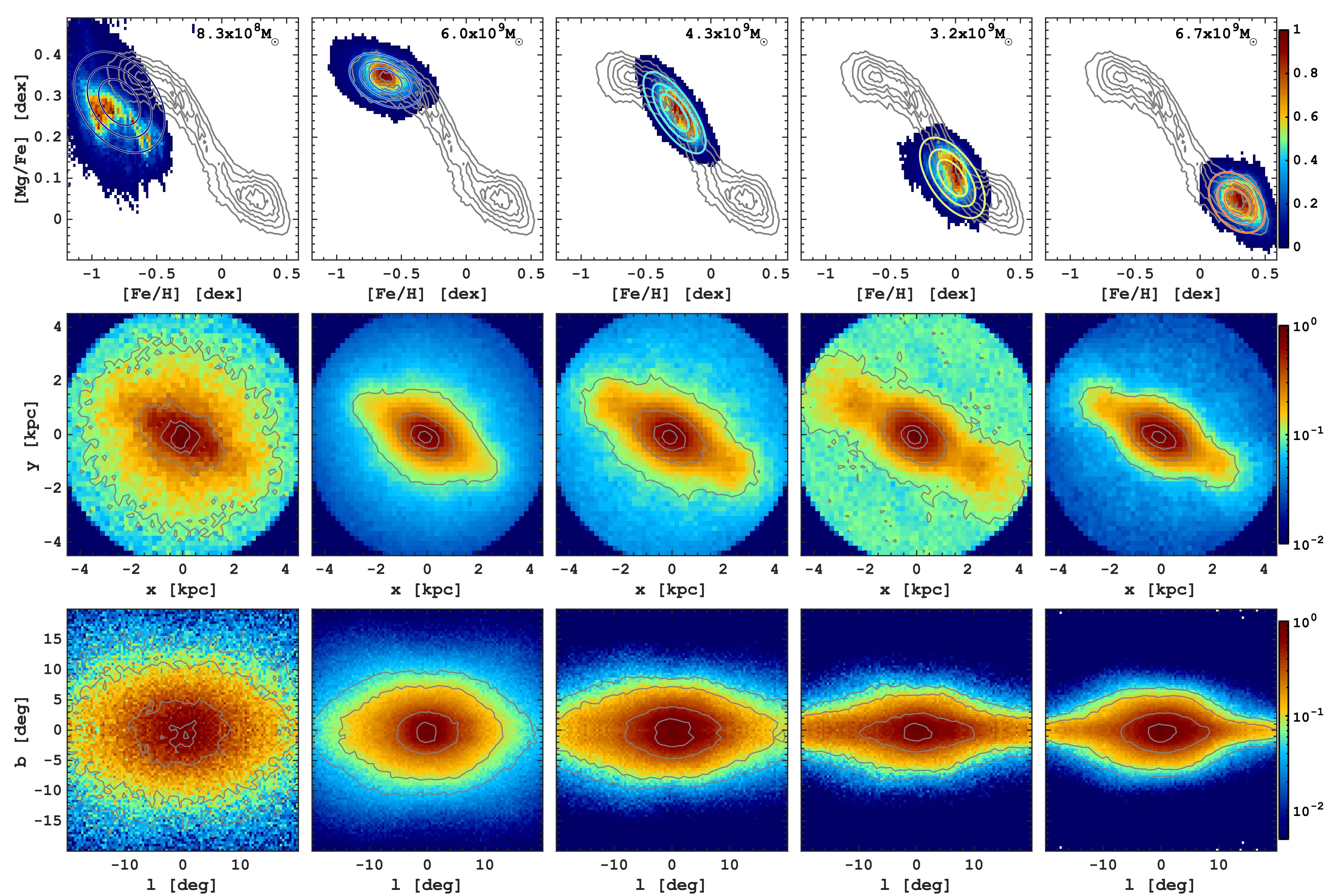}
    \caption{Density structure of the MW bulge components identified using 2D GMM in the \FeH-\MgFe plane. The top panels show the stellar mass density of each component obtained by multiplying the total stellar mass in the plane~(background contours) by the 2D GMM probability for a given component. The middle and bottom rows show the mass$\times$GMM-probability weighed density distribution in the XY and $(l,b)$ coordinates, respectively. The density is scaled by the maximum value in each panel.}
    \label{fig03::bulge_pops_dens}
\end{figure*}

\begin{figure*}
    \centering
    \includegraphics[width=1\hsize]{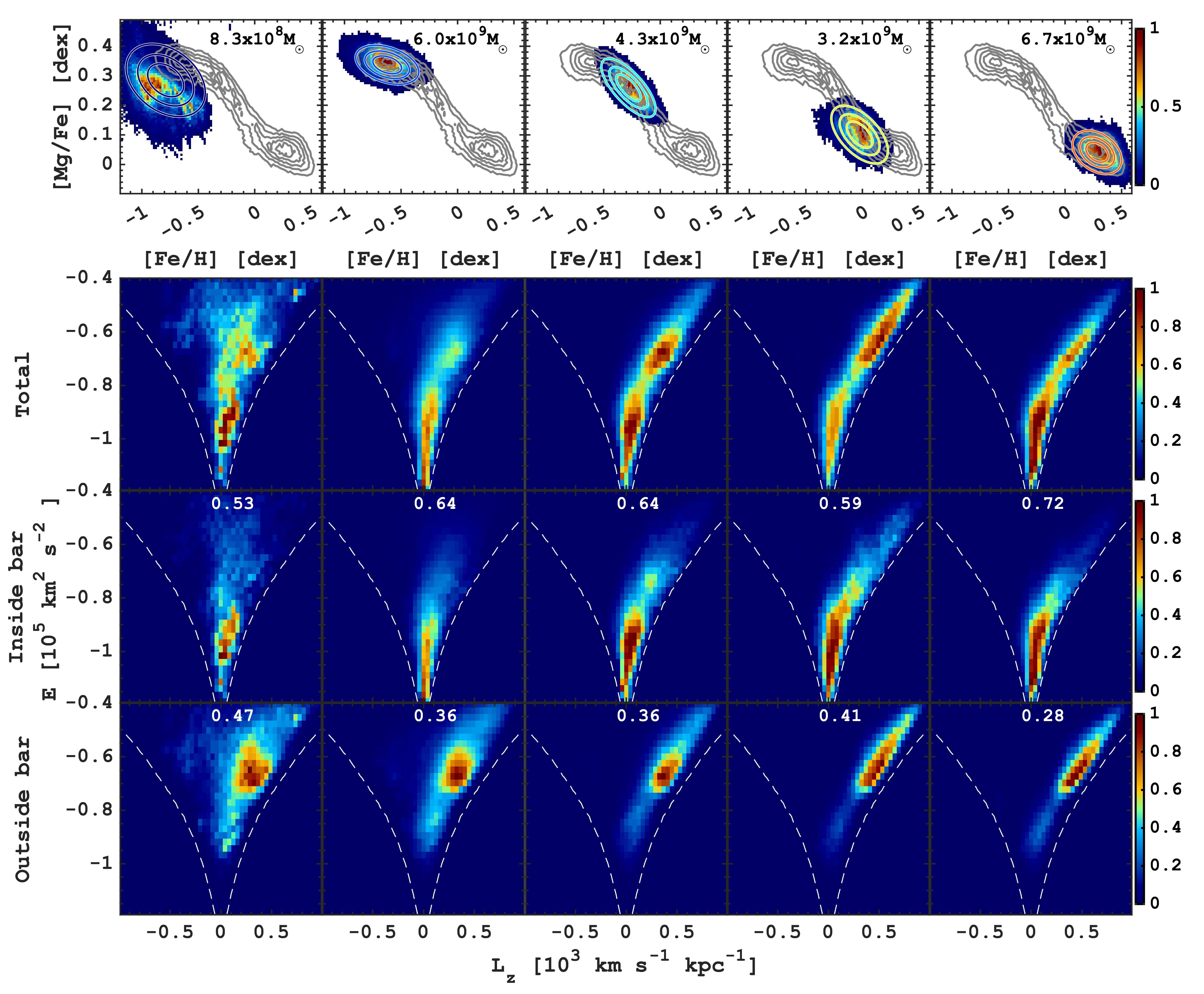}
    \caption{Energy-angular momentum structure of the MW bulge components identified using 2D GMM in the \FeH-\MgFe plane. The top panels are the same as in Fig.~\ref{fig03::bulge_pops_dens} and are given for reference. The bottom group of panels show the normalized stellar mass for the entire component~(top) and the region inside~(middle) and outside~(bottom) the bar. The numbers in the middle and bottom rows correspond to the mass fraction of a given component inside and outside the bar. The white dashed lines mark the minimum possible energy for a given value of the angular momentum.}
    \label{fig03::bulge_pops_elz}
\end{figure*}

To identify the centres of the GMM components in the \MgFe-\FeH plane, we used the entire bulge sample~(<3.5~kpc from the Galactic centre). Following previous studies~\citep{2017A&A...605A..89B, 2020MNRAS.499.1037R, 2021A&A...653A.143W}, we run the 2D GMM search for randomly sampled $10\%$ of the bulge populations. We repeat this procedure 50 times, and we adopt a full covariance GMM mode for each sample. We fitted the \MgFe-\FeH density distributions with an increasing number of Gaussian components from 2 to 6 and analysed the output using the Bayesian information criteria~\citep[BIC,][]{BIC}, which is minimal for five components, precisely the same number as in \cite{2013MNRAS.430..836N} or in \cite{2017A&A...605A..89B}. However, we underline that our GMM decomposition is different from the results of these studies. At this stage, we record only the positions~(\MgFe and \FeH) of the Gaussian components. The final centres of the 2D GMM components were obtained by averaging the positions obtained across all 50 random subsamples. Next, we run the GMM decomposition with the fixed centres of the components in different latitude slabs, allowing the model to find the independent orientation of each component at different positions relative to the midplane. 

The results of the 2D GMM decomposition of the \MgFe-\FeH relation for the bulge are presented in Fig.~\ref{fig03::bulge_gaussians2D}. The lines of different colours show the individual GMM components in \FeH-\MgFe plane~(left) and stacked \FeH~(middle) and \MgFe~(right) distributions. The fractional contributions of each component at different heights are marked with the same colour in the middle panels. 

The most natural explanation of the most metal-poor component~($\FeH \approx -0.8$~dex, dark blue) is directly related to its location in the \MgFe-\FeH plane, where chemically-defined halo and its ex-situ populations are located~\citep{2018Natur.563...85H, 2018ApJ...852...49H, 2019A&A...632A...4D, 2019MNRAS.482.3426M}. This is reinforced by its slightly lower \MgFe values compared to the high-$\alpha$ in-situ populations. The fractional contribution of the accreted populations increases from $5\%$ in the centre to $12\%$ at high latitudes. Interestingly, these values match very well various estimations of the mass of the stellar halo or spheroidal component in the MW bulge~\citep{2010ApJ...720L..72S, 2012AJ....143...57K, 2014A&A...567A.122D, 2016ApJ...821L..25K, 2017MNRAS.469.1587D}. 

The following GMM components essentially decompose the \FeH-\MgFe in-situ stars distribution onto four populations. The light blue and red correspond to the bulk of high- and low-$\alpha$ populations, respectively, while the intermediate~(cyan and green) two represent the tails of the bridge connecting the main components. Therefore, the physical meaning of the MW bulge GMM components is rather obvious. As demonstrated in the previous section, individual peaks of the MDF seen in our data do not correspond to specific orbital families, which have broad MDFs. Moreover, once we apply GMM decomposition, we do not recover these peaks, which seem to exist on top of the smooth distribution governed by the chemical evolution track(s) along metallicity. Essentially, the strength of the high- and low-$\alpha$ GMM components is defined by the masses of the thick and thin disc populations in the bulge region. The transition between these two populations, or the $\alpha$-knee, shapes the other two intermediate GMM components. The fact that we observe a gap in the \MgFe-\FeH relation around metallcity $\approx -0.05$~dex~(see Fig.~\ref{fig03::bulge_bimodality}, right panel) may imply the star formation quenching episode~\citep{2016A&A...589A..66H,2018A&A...618A..78H}. The manifestation of the quenching in the \MgFe-\FeH plane allows us to break down the bridge connecting the high- and low-$\alpha$ populations into two GMM components.

Although we do not suggest that the intermediate components~(cyan and green in Fig.~\ref{fig03::bulge_gaussians2D}) should be associated with distinct bulge populations, they can have very different spatial and kinematic characteristics, especially if the MW experienced rapid transformation at the corresponding epoch. To illustrate this in Fig.~\ref{fig03::bulge_pops_dens}, we present the face-on and $(l,b)$ stellar density projections of populations identified by the 2D GMM, which are obtained by multiplying the stellar density by the GMM component probability distributions. The maps show the trends presented in Fig.~\ref{fig03::bulge_density_feh}. The most metal-poor component (first column) appears as an elongated structure along the bar's major axis, with a weak flattening toward the midplane but lacks a prominent boxy or X-shaped morphology. It is worth noting that even when using the GMM probabilities as weights, some contamination from the high-$\alpha$ bulge population cannot be entirely ruled out because the membership for each component is based on the distance from the component centre. Hence, overlapping 2D distributions can not be fully separated unless more dimensions are used, which is the case of accreted/in-situ population distinction~\citep{2021MNRAS.508.1489F, 2022MNRAS.510.2407B, 2022ApJ...938...21M, 2023arXiv231005287K}. Nevertheless, the structure of this component is distinct from the next one (second column), which represents the bulk of the high-$\alpha$ populations. While the X-shaped structure is noticeable, it is not particularly sharp, giving the bar an overall oval-like appearance. In the third GMM component, the bar increases in size. Surprisingly, the longest bar is observed in the fourth GMM component, representing the low-\FeH tail of the low-$\alpha$ populations. This likely corresponds to the coldest stellar populations most tightly trapped by the bar. The bar remains relatively long for the most metal-rich component, though it is somewhat less extended than the previous component because of the rapidly decreasing scalelength of mono-abundance populations at $\FeH>+0.25$~(\citetalias{Mapping-disk}, \cite{2024arXiv240601706R}). 

The kinematics of the four individual in-situ components follow the trend we discussed above~(see Section.~\ref{sec3::bulge_kinematics}): increasing rotational support~(and decreasing velocity dispersion) with increasing metallicity~(see a relevant figure in Appendix~\ref{fig03::bulge_pops_kin}). The most metal-poor, potentially accreted GMM component still shows very weak rotation, which, as we already mentioned, can be contaminated by the metal-poor tail of the high-$\alpha$ populations.

Since the release of the second \Gaia catalogue, which provided phase-space data for millions of stars in the MW, numerous studies have sought to identify whether various stellar populations have in-situ or accreted origin using kinematic spaces~\citep[see][for compilation of the results]{2022ApJ...926..107M}. One of the most commonly used methods for identifying accreted systems is through the energy-angular momentum space~(\ELz), as these quantities are generally assumed to be conserved after a smaller galactic system falls into the MW~\citep{2000MNRAS.319..657H}. While this assumption is applicable for low-mass accreted systems, it is not entirely correct for massive galaxies, whose remnants are affected by dynamical friction~\citep{2017A&A...604A.106J, 2023A&A...673A..86P, 2024A&A...690A.136M}, and because the MW mass distribution~(stellar and DM) is non-axisymmetric and evolves over time~\citep{2021ApJ...920...10P, 2023A&A...677A..91K}. Although our orbit superposition approach is primarily focused on the main disc stellar populations and was not explicitly designed to accommodate accreted systems, it is worth understanding the structure of the in-situ component in \ELz space. This provides a reference framework, helping to distinguish accreted systems from genuine MW stars in future, especially in the innermost region of the bulge, where the remnants of anciently accreted systems should be buried.

In Figure~\ref{fig03::bulge_pops_elz}, we present the \ELz space for populations identified using a 2D GMM. The top row illustrates the GMM probability-weighted mass distributions in \ELz space, revealing a double-component structure across all populations, regardless of their metallicity. However, the specific details of these distributions vary slightly. To better understand the origin of this double-peak distribution, we separated the populations into two groups: stars located along the bar (within $\pm 1$ kpc of the bar major axis) and those outside the bar (within $<3.5$~kpc but not in the bar region). The bottom two rows in the figure demonstrate that bimodal \ELz distribution clearly arises from the superposition of bar and disc components. As expected, the bar component exhibits minimal rotation, as stars in this region tend to follow radial orbits, while the disc populations display weak but prograde rotation. Here, we underline that the non-rotating component, $\Lz \approx 0$, is not a signature of a spheroidal component but the bar.

This bimodal angular momentum distribution is not surprising and was also observed in an $N$-body simulation of a MW-like galaxy by \cite{2018A&A...616A.180F}. In their findings, thin disc stars were shown to lose more angular momentum due to bar/bulge formation, resulting in a more pronounced low-angular-momentum peak, whereas thick disc stars were less efficiently trapped by the bar and thus exhibited less pronounced bimodality in $L_z$. This trend is consistent with what we observe in Fig.~\ref{fig03::bulge_pops_elz}, where the fraction of low-\Lz stars increases with metallicity for each GMM-identified population, as indicated by the values displayed in each panel.

In most metal-poor \ELz bin, we observe several distinct features or clumps, which could potentially be remnants of yet unknown accreted galaxies, bulge/bar resonances~\citep{2024ApJ...971L...4D, 2024MNRAS.532.4389D}, or stars stripped from globular clusters in the bulge region~\citep{2023A&A...673A..44F}. Determining the true origin of these features requires further detailed investigation, and we do not assert their definitive nature at this stage. However, the fact that these features appear exclusively among the most metal-poor populations is in favour of the hypothesis that they are of ex-situ origin or related to old (not necessarily accreted) globular clusters.

\section{Summary}\label{sec3::summary}
Nowadays, the MW bulge is seen as the inner structure of the bar and is certainly a complex environment. It is a composite system with a mixture of different stellar populations, similar to other nearby extragalactic systems. Studying the bulge, which is a major component of our Galaxy, is an important step in clarifying how the MW formed and evolved. In this work, we applied our novel orbit superposition approach combined with the APOGEE DR 17 data to provide a unified present-day picture of the correlations between spatial structure, chemical abundances, kinematics, and orbital composition of stellar populations in the MW bulge. The main results of the work are summarized as follows.

\begin{itemize}

    \item We have shown that our orbit superposition method is precise enough, and the APOGEE giant stars sample provides sufficient data to successfully reconstruct the 3D stellar density structure of the MW bulge and bar region. This includes capturing the distinct X-shaped/boxy structure and the bimodal peanut density distribution observed along different lines of sight toward the bulge, further validating the effectiveness of our approach in recovering the complex structure of the inner MW.

    \item The fractional contributions of various orbital families that support the X-shaped boxy/peanut structure of the MW bulge highlight its intricate orbital dynamics, as predicted in secular formation scenarios. Our analysis reveals that banana-like orbits play a substantial role, contributing approximately $30\%$ to the overall bulge/bar structure. Meanwhile, the inner X-shaped orbits account for a more prominent $45\%$, underscoring their critical importance in shaping the distinctive morphology of the inner bulge.

    \item In the absence of any effects from the APOGEE footprint, we analysed the complete kinematics of the bulge. We are able to detect an increase~(decrease) of cylindrical rotation~(velocity dispersion) rate with increasing metallicity in the range from $-1.5$ to $+0.65$. Thanks to the high spatial coverage of the bulge, we detect small-scale asymmetries introduced by the bar, the major axis of which lies at an angle to the Sun - Galactic centre line. Additionally, we observed flattening in the 1D velocity dispersion profiles, characterized by shoulder-like features at various latitudes, which correspond to the X-shaped component of the bulge.
    
    \item Chemical abundance profiles vary greatly in the bulge region along longitude and latitude. 
    The radial metallicity profiles across the bulge region are not monotonic, while the vertical ones are negative but change with longitude. Close to the midplane~($|b|<2^\circ$), the mean metallicity decreases towards the Galactic centre, suggesting the lower metallicity of the MW bulge compared to the surrounding disc -- metal-poor bulge in the $(l,b)$ projection. 
    This picture is inverted already at latitudes $b>2^\circ$. Once we measure the mean metallicity, including high distances from the midplane, we find that the inner bulge region and the bar major axis are more metal-rich compared to the surrounding disc. Therefore, while the mean metallicity of the bulge is slightly subsolar, we argue that it is metal-rich, which is in perfect agreement with nearby boxy/peanut barred galaxies and simulations. 
    
    \item Overall, there is no universal metallicity gradient value that can characterise the MW bulge. The radial gradients very closely trace the X-shaped bulge density structure, reaching the maximum of $\rm 0.01~dex~deg^{-1}$ near the tips of the X-shaped component. The spatial variations of the vertical gradient better trace the boxy component, reaching the maximum values~($\rm -0.07~dex~deg^{-1}$) at high latitudes, on the edge of the bulge, where the transition from low- to high-$\alpha$ dominated region is very prominent. The vertical metallicity gradient reaches the smallest values in the innermost region~($|l|<3^\circ$ and $|b|<2^\circ$), where the metallicity profile is flat. 
    
    \item We showed that the MW bulge is composed of two primary components: high- and low-$\alpha$ populations. These correspond to the inner thick disc, which is metal-poor with subsolar metallicity, and the thin disc, which is metal-rich with supersolar metallicity, respectively. The mass ratio between these chemically distinct populations is 4:3, indicating a slight predominance of the metal-poor component over the metal-rich one.
    
    \item The metallicity distribution function of the MW bulge is broad, featuring two dominant peaks at approximately $-0.7$~dex and $+0.3$~dex, which correspond to the maximum density of the high- and low-$\alpha$ populations. Between these two main components, we observe multiple narrow peaks, which are present in the orbital families that support the bulge but absent in the MDF of the background disc populations. However, we do not propose a direct one-to-one connection between individual MDF peaks and specific orbital families, as the latter exhibit a range of metallicities and have relatively broad MDFs themselves. The latter illustrates that stars formed at different epochs can be trapped by the X-shaped/boxy structure with varying efficiency, depending on the initial distribution function of these populations and the evolution of the bar/bulge parameters. 
    
    \item Although the MW bulge is primarily dominated by two main components, its broad MDF cannot be fully explained by these components alone. To delve into the complexity of the bulge chemical abundance composition, we decomposed it using the 2D GMM approach in the stellar-mass weighed \FeH-\MgFe plane. We find that the optimal number of 2D GMM components is five, where the most metal-poor one is dominated by stars with a chemical composition typical for accreted populations. Although our orbit superposition approach is not tailored to include distinct halo populations, we estimate that the mass of stars with ex-situ chemical composition -- those accreted from outside the MW -- amounts to approximately $8 \times 10^8 \Msun$, for stars with metallicity above $\approx -1.2$ dex. Between the primary high- and low-$\alpha$ populations, we identified two additional 2D GMM components. These metal-intermediate components provide a more detailed breakdown of the $\alpha$-knee and help to characterize the 'bridge' that connects the high- and low-$\alpha$ populations. We, therefore, suggest that these intermediate MDF components do not correspond to distinct bulge populations but rather describe the transition between the two main chemical components.
    
\end{itemize}

In this study, we provided a comprehensive view of the present-day MW bulge. By analyzing its chemo-kinematic properties, we confirm that the bulge is a product of vertical bar instability, which captured and reshaped the inner populations of both the thin and thick discs over time. Aside from a relatively small fraction of stars from accreted populations, we find no significant evidence of a spheroidal component in the bulge. This suggests that the MW bulge shares many characteristics with secularly formed bulges in other barred galaxies, reinforcing the idea that the MW is a typical barred galaxy.

A few pieces still remain unresolved in the MW bugle story, particularly regarding the precise mechanism behind the bulge formation. It is unclear whether it was shaped by vertically asymmetric buckling~\citep{1991A&A...252...75P, 1994ApJ...425..551M} or a more symmetric, slower resonant heating process relative to the midplane~\citep{2002AJ....124..722Q, 2014MNRAS.437.1284Q}. Additionally, the exact timing of the X-shaped structure’s formation remains unconstrained. Although $N$-body simulations indicate that vertical instability often occurs relatively quickly after the bar's development, no clear chemo-kinematic indicators have yet been identified in models allowing us to constrain the epoch of bulge formation in the MW~\citep[see the discussion in][]{2019MNRAS.485.5073D}. Further theoretical efforts are required to address these open questions. On the observational side, more data will be provided in the inner parts of the MW by forthcoming spectroscopic surveys like 4MOST~\citep{2019Msngr.175....3D}, SDSS-V~\citep{2023ApJS..267...44A}, and MOONS~\citep{2020Msngr.180...18G}. Additionally, the continued application of the orbit superposition approach and, more generally, Schwarzschild methods will be crucial in advancing our understanding of the MW bulge and its formation history.

\begin{acknowledgements}
SK expresses gratitude to Ortwin Gerhard for an insightful discussion of relevant topics in April 2024. \\

\thanksgaia \\

\thankssdss

\end{acknowledgements}

\bibliographystyle{aa}
\bibliography{refs}

\begin{appendix}
\section{Extra figures}

\begin{figure}
    \centering
    \includegraphics[width=1\hsize]{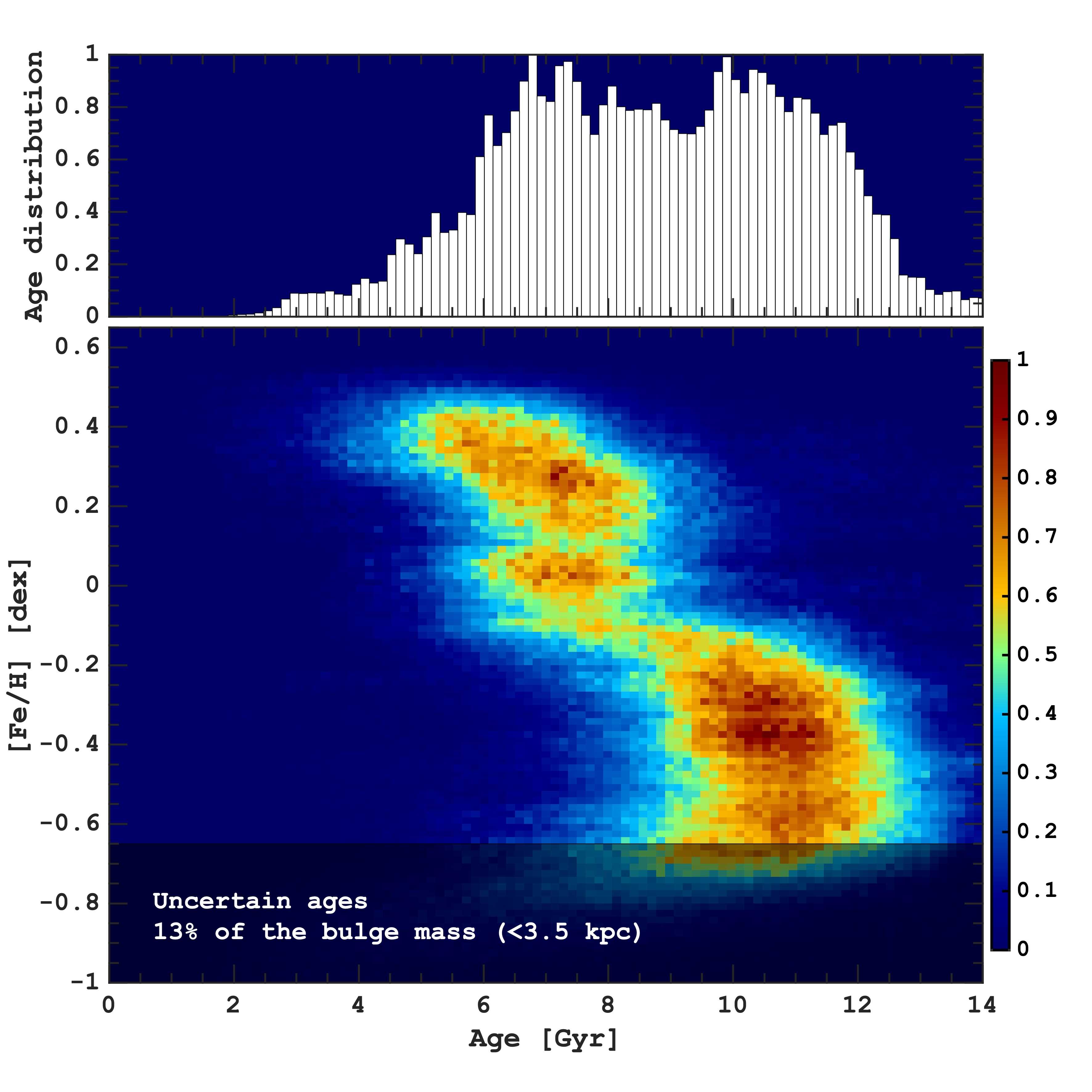}
    \caption{Stellar mass-weighted age distribution~(top) and age-metallicity relation~(bottom) in the MW bulge region~($<3.5$~kpc from the Galactic centre). The metallicity range~($\FeH<-0.65$~dex, \cite{2024AJ....167...73S}) with uncertain ages is shaded in the bottom panel and not included in the top distribution, accounting for about $13\%$ of the bulge mass.}
    \label{fig03::bulge_AMR}
\end{figure}



\begin{figure*}
    \centering
    \includegraphics[width=1\hsize]{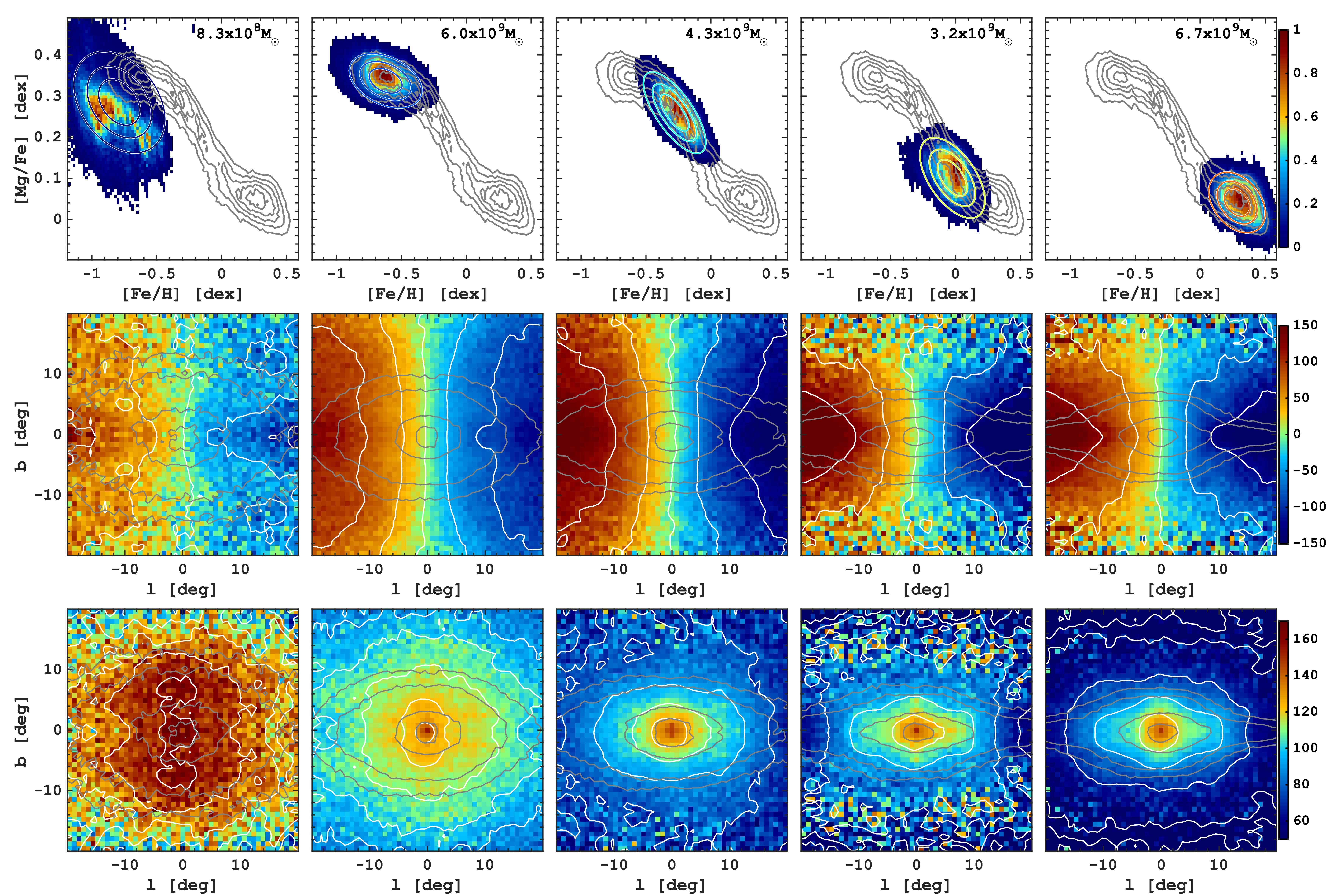}
    \caption{Kinematics of the MW bulge populations identified using 2D GMM in the \FeH-\MgFe plane. The top panels are the same as in Fig.~\ref{fig03::bulge_pops_dens} and are given for reference. The middle and bottom panels show each component's line-of-sight velocity and dispersion maps, weighted by the mass multiplied by the GMM probabilities. The grey contours show corresponding density distribution, while the white ones highlight the velocity/dispersion structure.
    }
    \label{fig03::bulge_pops_kin}
\end{figure*}

\end{appendix}

\end{document}

%% file: input.tex
\usepackage[breaklinks=true,colorlinks,citecolor=blue]{hyperref}
\usepackage{graphicx,animate}
\usepackage{enumerate}
\usepackage{epsfig}
\usepackage{color}
\usepackage{txfonts}
\usepackage{natbib}
\usepackage{multirow}
\usepackage{ulem} 
\usepackage{amssymb}

\usepackage{movie15}

\usepackage{booktabs}
\usepackage{array}   

\usepackage{academicons}

\newcommand{\orcid}[1]{\href{https://orcid.org/#1}{\textcolor[HTML]{A6CE39}{\aiOrcid}}}
\usepackage{orcidlink}

\usepackage{xcolor}

\defcitealias{Mapping-model}{Paper I}
\defcitealias{Mapping-disk}{Paper II}
\defcitealias{Mapping-bulge}{Paper III}
\defcitealias{Mapping-extragalactic}{Paper IV}
\defcitealias{Mapping-sfh}{Paper V}

\definecolor{ao}{rgb}{0.0, 0.5, 0.0}

\newcommand{\kmps}{\rm km~s\ensuremath{^{-1} }\,}

\newcommand{\Msun}{M\ensuremath{_\odot}}

\newcommand{\Gaia}{{\it Gaia}\,}

\newcommand{\Lz}{\ensuremath{\rm L_z}\,}
\newcommand{\ELz}{\ensuremath{\rm E-L_z}\,}

\newcommand{\MgFe}{\ensuremath{\rm [Mg/Fe]}\,}
\newcommand{\FeH}{\ensuremath{\rm [Fe/H]}\,}

\newcommand{\papername}{Rediscovering the Milky Way with orbit superposition approach and APOGEE data}

\newcommand{\thankssdss}{Funding for the Sloan Digital Sky Survey IV has been provided by the Alfred P. Sloan Foundation, the U.S. Department of Energy Office of Science, and the Participating Institutions. SDSS acknowledges support and resources from the Center for High-Performance Computing at the University of Utah. The SDSS web site is www.sdss4.org. \\

SDSS is managed by the Astrophysical Research Consortium for the Participating Institutions of the SDSS Collaboration including the Brazilian Participation Group, the Carnegie Institution for Science, Carnegie Mellon University, Center for Astrophysics | Harvard \& Smithsonian (CfA), the Chilean Participation Group, the French Participation Group, Instituto de Astrofísica de Canarias, The Johns Hopkins University, Kavli Institute for the Physics and Mathematics of the Universe (IPMU) / University of Tokyo, the Korean Participation Group, Lawrence Berkeley National Laboratory, Leibniz Institut für Astrophysik Potsdam (AIP), Max-Planck-Institut für Astronomie (MPIA Heidelberg), Max-Planck-Institut für Astrophysik (MPA Garching), Max-Planck-Institut für Extraterrestrische Physik (MPE), National Astronomical Observatories of China, New Mexico State University, New York University, University of Notre Dame, Observatório Nacional / MCTI, The Ohio State University, Pennsylvania State University, Shanghai Astronomical Observatory, United Kingdom Participation Group, Universidad Nacional Autónoma de México, University of Arizona, University of Colorado Boulder, University of Oxford, University of Portsmouth, University of Utah, University of Virginia, University of Washington, University of Wisconsin, Vanderbilt University, and Yale University.}

\newcommand{\thanksgaia}{This work presents results from the European Space Agency (ESA) space mission Gaia. Gaia data are being processed by the Gaia Data Processing and Analysis Consortium (DPAC). Funding for the DPAC is provided by national institutions, in particular the institutions participating in the Gaia Multi-Lateral Agreement (MLA). The Gaia mission website is https://www.cosmos.esa.int/gaia. The Gaia Archive website is http://archives.esac.esa.int/gaia.}

\defcitealias{Mapping-1}{Paper I}
\defcitealias{Mapping-2}{Paper II}
\defcitealias{Mapping-3}{Paper III}
\defcitealias{Mapping-4}{Paper IV}
\defcitealias{Mapping-5}{Paper V}